\documentstyle[12pt,epsfig]{article}
\newcommand{\n}{\hspace*{-2.5mm}}
\newcommand{\simgt}{\,\rlap{\lower 3.5 pt \hbox{$\mathchar \sim$}} \raise 1pt
 \hbox {$>$}\,}
\newcommand{\simlt}{\,\rlap{\lower 3.5 pt \hbox{$\mathchar \sim$}} \raise 1pt
 \hbox {$<$}\,}

\catcode`@=11
\newcount\@tempcntc
\def\@citex[#1]#2{\if@filesw\immediate\write\@auxout{\string\citation{#2}}\fi
  \@tempcnta\z@\@tempcntb\m@ne\def\@citea{}\@cite{\@for\@citeb:=#2\do
    {\@ifundefined
       {b@\@citeb}{\@citeo\@tempcntb\m@ne\@citea\def\@citea{,}{\bf ?}\@warning
       {Citation `\@citeb' on page \thepage \space undefined}}%
    {\setbox\z@\hbox{\global\@tempcntc0\csname b@\@citeb\endcsname\relax}%
     \ifnum\@tempcntc=\z@ \@citeo\@tempcntb\m@ne
       \@citea\def\@citea{,}\hbox{\csname b@\@citeb\endcsname}%
     \else
      \advance\@tempcntb\@ne
      \ifnum\@tempcntb=\@tempcntc
      \else\advance\@tempcntb\m@ne\@citeo
      \@tempcnta\@tempcntc\@tempcntb\@tempcntc\fi\fi}}\@citeo}{#1}}
\def\@citeo{\ifnum\@tempcnta>\@tempcntb\else\@citea\def\@citea{,}%
  \ifnum\@tempcnta=\@tempcntb\the\@tempcnta\else
   {\advance\@tempcnta\@ne\ifnum\@tempcnta=\@tempcntb \else \def\@citea{--}\fi
    \advance\@tempcnta\m@ne\the\@tempcnta\@citea\the\@tempcntb}\fi\fi}
\catcode`@=12
\begin{document}
\title{\vskip-3cm{\baselineskip14pt
\centerline{\normalsize DESY 97--110\hfill ISSN~0418--9833}
\centerline{\normalsize MPI/PhT/97--001\hfill}
\centerline{\normalsize hep--ph/9706369\hfill}
}
\vskip1.5cm
Charmonium Production via Fragmentation at DESY HERA}
\author{{\sc Bernd A. Kniehl$^1$ and Gustav Kramer$^2$}\\
$^1$ Max-Planck-Institut f\"ur Physik (Werner-Heisenberg-Institut),\\
F\"ohringer Ring 6, 80805 Munich, Germany\\
$^2$ II. Institut f\"ur Theoretische Physik\thanks{Supported
by Bundesministerium f\"ur Forschung und Technologie, Bonn, Germany,
under Contract 05~7~HH~92P~(5),
and by EU Program {\it Human Capital and Mobility} through Network
{\it Physics at High Energy Colliders} under Contract
CHRX--CT93--0357 (DG12 COMA).},
Universit\"at Hamburg,\\
Luruper Chaussee 149, 22761 Hamburg, Germany}
\date{}
\maketitle
\begin{abstract}
The cross section for the photoproduction of large-$p_T$ $J/\psi$ mesons
at HERA is calculated at next-to-leading order,
adopting a perturbative approach to describe the fragmentation 
of charm quarks and gluons into $J/\psi$ mesons. We treat the charm quark
according to the massless factorization scheme, where it is assumed
to be one of the active flavours inside the proton and the resolved photon.
We present inclusive distributions in transverse momentum and rapidity,
including the contributions due to direct and resolved photons.
The importance of the colour-octet components of the $J/\psi$ wave function,
which contribute to the fragmentation process, is emphasized.
In addition to prompt $J/\psi$ production, we consider also the production of
$\chi_{cJ}$ states followed by radiative decays to $J/\psi$ mesons, both in
the colour-singlet and colour-octet channels.

\end{abstract}
\newpage

\section{Introduction}

The production of heavy quarkonia in high-energy collision
has conventionally been described within the colour-singlet model
(CSM) \cite{ber}. In this model, a non-relativistic approximation
is used to describe the binding of the heavy-quark pair,
produced via parton-fusion processes,
into a quarkonium state. The heavy-quark pair is
projected onto a physical quarkonium state using a
colour-singlet projection and an appropriate spin projection.
Experimentally, $J/\psi$ production is measured most easily
through the detection of the leptonic decays. Then, the inclusive $J/\psi$
yield consists of the prompt $J/\psi$ production and the production of the
$c\bar c$ ${}^3\!P_J$ states $\chi_{cJ}$ $(J=0,1,2)$, which decay
electromagnetically into the lower-lying $J/\psi$ state.
The latter must also be taken into
account if one wishes to compare theoretical predictions with experiments
where $\chi_{cJ}$ production could not be measured separately.
The CSM has been successfully applied \cite{bai} to
explain large-$p_T$ $J/\psi$ production at the relatively
low energies ($\sqrt s=63$~GeV) of the CERN ISR experiments.
After the advent of the CERN $Sp\bar pS$ collider with $\sqrt s=630$~GeV,
it was found \cite{glo} that also the production of bottom quarks and their
subsequent weak decay into $J/\psi$ mesons is an important source of
$J/\psi$ production. In the UA1 experiment, it was not
possible to separate the bottom-quark contribution.

Later, the CDF and D0 collaborations, when measuring
$J/\psi$ production at the Fermilab Tevatron ($\sqrt s=1.8$~TeV),
were able to subtract the bottom-quark contribution from
the total yield.
Since also the yield of the $\chi_{cJ}$ mesons decaying into $J/\psi$ mesons
could be separated, a measurement of prompt $J/\psi$ production at large $p_T$
was possible.
(We shall use the term {\it prompt} instead of {\it direct} to denote all
$J/\psi$ mesons not originating from decays, {\it i.e.}, $B$-meson or
$\chi_{cJ}$ decays.)
The cross section of prompt inclusive $J/\psi$ production measured by CDF and
D0 \cite{abe} turned out to be an order of magnitude larger than the
prediction of the CSM.
It is well known that, at large $p_T$, the dominant mechanism of
light-quarkonium production is via the fragmentation of gluons and quarks,
while the parton-fusion process is suppressed by powers $m^2/p_T^2$, where $m$
is the mass of the quarks bound in the quarkonium state.
So, the fusion process, which is a higher-twist contribution,
can contribute significantly only at extremely small $p_T$, which for the
light quarkonia is in the soft region outside the perturbative regime.
For $J/\psi$ production, this region, due to the
larger charm-quark mass $m_c$, extends to larger $p_T$.
But eventually, at $p_T\gg m_c$, the fragmentation process, although of higher
order in $\alpha_s$ compared to the fusion process, can exceed the fusion
process.
In fact, it was pointed out by Braaten and Yuan \cite{bra}
that, in addition to the parton-fusion contributions described
by the CSM, fragmentation of gluons and charm quarks
is an important source for large-$p_T$ mesons in hadron-hadron collisions.

$J/\psi$ production in high-energy $ep$ collisions at HERA
is dominated by photoproduction, where the electron
is scattered by a small angle producing photons of
almost zero virtuality.
First experimental results in the kinematical region 30~GeV$<W<150$~GeV, where
$W$ is the $\gamma p$ centre-of-mass (CM) energy, have been presented by the
H1 collaboration \cite{aid}.
In this work, the inelastic process is analysed for the
first time. Most of the events come from the
elastic and diffractive processes. They can be
separated by measuring the $J/\psi$ energy spectrum
in terms of the inelasticity variable
$\tilde z=p_p\cdot p_{J/\psi}/p_p\cdot p_\gamma$,
with $p_p$, $p_\gamma$, and $p_{J/\psi}$ being the four-momenta of the proton,
the photon, and the $J/\psi$ meson, respectively. For elastic/diffractive
events, one has $\tilde z\approx1$. A clean sample of inelastic
events can be obtained by selecting events with $\tilde z<0.9$.
Due to the limited integrated luminosity in the 1994 HERA
operation, the inelastic $J/\psi$ production cross section $d\sigma/dp_T^2$
could only be measured for $p_T^2<6$~GeV${}^2$ \cite{aid}. In this range of
$p_T^2$, the inelastic production proceeds predominantly
through the fusion mechanism in the CSM. In fact,
the experimental $p_T$ and $\tilde z$ distributions are very well described,
as for both shape and absolute normalization, by the CSM
next-to-leading-order (NLO) calculation of Kr\"amer \cite{kra},
using up-to-date NLO parton density functions (PDF's) of the proton.

As is well known, high-energy photoproduction proceeds
through two distinct mechanisms: direct photoproduction,
where the photon couples directly to the quarks, and resolved
photoproduction, where the photon interacts through one of
its partonic components, quarks or gluons, via hard scattering with the
partonic constituents of the proton.
For $J/\psi$ photoproduction within the CSM, the resolved-photon process is
\begin{equation}
\gamma + p\to g_\gamma+g_p\to c\bar c[\,\underline{1},{}^3\!S_1]+g\to J/\psi+g.
\end{equation}
This means that the gluon coming from the proton interacts with the gluon from 
the proton to create a ${}^3\!S_1$ colour-singlet $c\bar c$ pair, which 
subsequently fuses into a $J/\psi$ meson.
This process has been analysed \cite{jun}
with the result that its contribution to the overall
cross section in the CSM is small everywhere,
except in the very-low-$\tilde z$ region. Contributions
from the production and radiative decay of $\chi_{cJ}$ states are
of similar magnitude; see Ref.~\cite{jun} for a review of these
and other resolved-photoproduction mechanisms within the CSM.

Within the factorization approach, $J/\psi$ production can
also proceed through colour-octet $c\bar c$ pairs. They appear
for direct photoproduction as well as for resolved
photoproduction. Colour-octet configurations which
contribute to inelastic direct $J/\psi$ photoproduction are
produced through the following subprocesses:
\begin{eqnarray}
\gamma+g&\n\to\n&c\bar c[\,\underline{8},{}^1\!S_0]+g,\nonumber\\
\gamma+g&\n\to\n&c\bar c[\,\underline{8},{}^3\!S_1]+g,\nonumber\\
\gamma+g&\n\to\n&c\bar c[\,\underline{8},{}^3\!P_J]+g,
\end{eqnarray}
where $J=0,1,2$.
Other contributions, such as light-quark initiated ones, are
strongly suppressed at HERA energies and can be safely
neglected. Adopting the colour-octet matrix elements
obtained through fits to prompt $J/\psi$ data from the Tevatron
\cite{cho}, one finds that colour-singlet and colour-octet
contributions to inelastic direct photoproduction are of
comparable size \cite{cac}.
In the case of resolved photoproduction, the following colour-octet processes
contribute:
\begin{eqnarray}
g_\gamma+g_p&\n\to\n&c\bar c[\,\underline{8},{}^1\!S_0]+g,\nonumber\\
g_\gamma+g_p&\n\to\n&c\bar c[\,\underline{8},{}^3\!S_1]+g,\nonumber\\
g_\gamma+g_p&\n\to\n&c\bar c[\,\underline{8},{}^3\!P_J]+g.
\end{eqnarray}
These are the same processes that contribute significantly to $J/\psi$
production in $p\bar p$ collisions at the Tevatron \cite{cho}.
Their contribution to inelastic $J/\psi$ photoproduction has also been
calculated \cite{mca} and found to be non-negligible.
In fact, it is comparable to the cross section of colour-singlet direct
photoproduction \cite{mca}. The distributions in the
inelasticity variable $\tilde z$ are, however, quite different.
The resolved-photon cross section (colour-singlet plus colour-octet)
is dominant at small $\tilde z$, whereas the direct-photon
contribution is large towards $\tilde z=1$.

For $\chi_{cJ}$ production, the colour-octet channels are
much more important. In the case of direct
photoproduction, the production of colour-singlet $\chi_{cJ}$
states via
\begin{equation}
\gamma + g\to c\bar c[\,\underline{1},{}^3\!P_J]+g\to\chi_{cJ}+g
\end{equation}
is forbidden to leading order in $\alpha_s$, so that the
colour-octet processes
\begin{eqnarray}
\label{cfo}
\gamma+g&\n\to\n&c\bar c[\,\underline{8},{}^3\!S_1]+g,\nonumber\\
\gamma+q(\bar q)&\n\to\n&c\bar c[\,\underline{8},{}^3\!S_1]+q(\bar q),
\end{eqnarray}
where the colour-octet ${}^3\!S_1$ $c\bar c$ pair fuses into a
physical $\chi_{cJ}$ particle, are the dominant ones. Nevertheless,
using the matrix elements obtained through fits to the Tevatron data
\cite{cho,ebr,gre},
the contribution of the channels (\ref{cfo}) is still
small compared to the dominant CSM contribution to $J/\psi$
production \cite{mca}. The cross sections for resolved photoproduction
of $\chi_{cJ}$ mesons in the colour-octet channels have not yet been 
estimated.

All these results refer to $J/\psi$ production at $p_T$ values of order $m_c$,
where we may expect the higher-twist contribution to be dominant.
At sufficiently large $p_T$, the fragmentation process is supposed
to be the dominant mechanism of $J/\psi$ photoproduction.
First estimates of the fragmentation contribution to $J/\psi$
photoproduction have been presented in Ref.~\cite{god};
see also Ref.~\cite{sal} for earlier work within the CSM.
These authors considered, at leading order (LO), the fragmentation of charm
quarks and gluons produced through the direct-photon processes
$\gamma+q\to g+q$ and $\gamma+g\to c+\bar c$.
They did not take into account resolved photoproduction, which is certainly
important, too.
In the case of $J/\psi$ production in $p\bar p$ collisions at the Tevatron,
it is the analogous mechanism which makes the dominant contribution
\cite{ebr,gre}.

It is the purpose of this work to study $J/\psi$ photoproduction at large
$p_T$ including all direct- and resolved-photoproduction channels at NLO.
We shall focus our attention on the production via fragmentation.
We shall consider the fusion mechanism only for comparison;
for simplicity, we shall only include its dominant LO contribution, which
arises from direct photoproduction in the CSM.
The superposition of the NLO cross sections for the fusion and
fragmentation mechanisms is left to future work.

The framework for our calculation of the NLO fragmentation cross section
is similar to the one employed recently for the calculation of inclusive
$D^{*\pm}$ photoproduction \cite{kni}. At large $p_T$, one may neglect the mass
of the charm quark. The collinear singularities corresponding
to the $\alpha_s\ln(p_T^2/m_c^2)$ terms of the massive-charm scheme, which is
used, for example, for predictions at $p_T\simlt m_c$, are then absorbed into
the charm-quark PDF's and the fragmentation functions
(FF's) in the same way as for the lighter $u$, $d$, and $s$ quarks.
The FF's for the fragmentation of charm quarks and gluons into $J/\psi$ mesons
must be known for this approach at the scale of production, which is usually
taken to be $\mu=m_T$, where $m_T=\sqrt{M_{J/\psi}^2+p_T^2}$ is the transverse
mass of the $J/\psi$ meson with mass $M_{J/\psi}\approx2m_c$.
These FF's can be computed at the starting scale $\mu_0=M_{J/\psi}$
in nonrelativistic QCD adopting as nonperturbative input the colour-singlet and
colour-octet transition-matrix elements, which also enter the fusion approach.
At larger scales $\mu\gg M_{J/\psi}$, the large
logarithms in $\mu/M_{J/\psi}$ are resummed using the usual Altarelli-Parisi
evolution equations.
At the starting scale, only the fragmentation of gluons and charm quarks
need to be included, while the light-quark contributions are negligible.
The latter will be generated at higher scales via evolution.

The outline of our work is as follows. In Section~2,
we shall shortly describe the formalism for the perturbative FF's
and discuss the transition from massless to massive factorization
used also in our previous work \cite{kni}. In Section~3, we shall present
our NLO predictions for the various fragmentation contributions to the cross
section of $J/\psi$ photoproduction in $ep$ collisions at HERA
as functions of $p_T$ and rapidity $y$, and compare them with the LO
CSM results based on the $c\bar c$ fusion mechanism.
Specifically, we shall distinguish between direct-photon and resolved-photon,
prompt and non-prompt (only via $\chi_{cJ}$ states), colour-singlet and
colour-octet fragmentation production.
Our conclusions will be summarized in Section~4.

\section{Charmonium fragmentation functions}

In this section, we shall describe the assumptions underlying
the massless approach, which is very similar to that
in our work on inclusive $D^*$ production \cite{kni}.
For the reader's convenience, we shall repeat here the basic ideas developed
in Ref.~\cite{kni}.
As already mentioned in the Introduction, two mechanism contribute to the
photoproduction of charm quarks in $ep$ collisions:
$(i)$ In the direct photoproduction mechanism, the photon couples directly
to the quarks, which besides the massless $u$, $d$, and $s$
quarks, also include the massless charm quark.
In this case, at least in the LO processes,
no spectator particles travel along the momentum direction of the photon.
$(ii)$ In the resolved photoproduction mechanism, the photon splits
up into fluxes of $u$, $d$, $s$, $c$ quarks and gluons, which
then interact with the partons coming from the proton leading
to the production of quarks, including charm quarks,
and gluons at large $p_T$. The contributing parton-level
processes are the same as in the case of $J/\psi$ production
in hadron-hadron collisions. The quarks and gluons
coming out of the photon are accompanied by a
spectator jet, which also travels in the photon direction. Therefore,
the $\gamma p$ cross section depends not only on the PDF's of
the proton but also on those of the photon. The main
difference with respect to the usually considered massive-charm
scheme, which to our knowledge has not yet been worked out
for $J/\psi$ production via fragmentation, is that the charm quark also
contributes via the PDF's of the proton and the photon, {\it i.e.}, charm
is already an active flavour in the initial state. Compared
to $D^{*\pm}$ production, where the $D^{*\pm}$ mesons are almost exclusively
produced via charm-quark fragmentation, in $J/\psi$ production the
fragmentation process $g\to J/\psi+X$ is very important as well, as we shall
see in the next section. The massless-charm approach is
justified if a large scale is governing the production process.
In our case, this is the transverse momentum of the $J/\psi$ meson, with
$p_T\gg M_{J/\psi}$.
In this region, non-singular mass terms are suppressed
in the cross section by powers of $m_c/p_T$. The
important mass terms appear if the virtuality of the charm quark
is small. This occurs in the initial state if the charm quark is
emitted from the proton or photon, and in the final state if
the partons emit a $c\bar c$ pair. These two contributions lead
to the $\alpha_s\ln(p_T^2/m_c^2)$ terms in the massive scheme. In our massless
approach, they are summed into the scale-dependent proton and photon PDF's
and $J/\psi$ FF's.

To obtain the final $J/\psi$ production cross sections in NLO,
the following steps are taken. $(i)$ The NLO hard-scattering
cross sections for the direct- and resolved-photon processes
are calculated in the massless approximation with $n_f=4$ active
flavours \cite{aur}. The collinear singularities are subtracted according
to the $\overline{\mbox{MS}}$ scheme. Since the charm quark is taken
to be massless, the singularities from its splittings are
subtracted as well. $(ii)$ The charm quark is accommodated in the
PDF's of the proton and photon as a light flavour.
The finite mass of the charm quark is taken into account by
including it in the evolution of the PDF's in such a way
that these are only non-vanishing above a scale set
by its mass. $(iii)$ The FF's characterize the hadronization
of the massless partons, including the charm quark, into $c\bar c$
bound states, {\it i.e.}, the $J/\psi$ and $\chi_{cJ}$ mesons.
Similarly to the
fragmentation into light mesons, these FF's are non-perturbative
input and must be determined by experiment, for example from
the cross section of inclusive $J/\psi$ and $\chi_{cJ}$ production in
hadron-hadron collisions. This information determines
the FF's at some starting scale, $\mu_0$.

An alternative methods is to calculate, within perturbative QCD, universal
starting conditions for the FF's at a scale $\mu_0$ of order $M_{J/\psi}$.
This allows one to include in a straightforward manner
the colour-octet contributions with parameters
related to the ones obtained from the analysis of
colour-octet fusion.

The structure of the perturbative FF's is based on the general factorization
analysis of the production and decay of
heavy quarkonia by Bodwin, Braaten, and Lepage \cite{bod}.
This factorization formalism allows for the FF's of
charmonium production to be factorized into short-distance
coefficients that describe the
production rate of a $c\bar c$ pair within a region of
size $1/m_c$, and long-distance factors that contain the 
non-perturbative dynamics responsible for the formation of the bound
state $H$ containing the $c\bar c$ pair. The FF, at factorization scale $\mu$,
for a parton $a$ to fragment into the charmonium state $H$ with longitudinal
momentum fraction $z$ can be written as
\begin{equation}
\label{dec}
D_a^H(z,\mu)=\sum_nd_a^n(z,\mu)\langle0|{\cal O}_n^H|0\rangle,
\end{equation}
where $n$ labels different colour components (singlet or octet) and spin
quantum numbers to be specified later.
${\cal O}_n^H$ are the local four-fermion operators defined in
Ref.~\cite{bod} in terms of the fields of nonrelativistic QCD.
The short-distance coefficients $d_a^n(z,\mu)$ can be computed at the
starting scale $\mu_0$ using perturbation theory in $\alpha_s(\mu_0)$.
The dependence on the observed charmonium state $H$ only appears in the
long-distance factors $\langle0|{\cal O}_n^H|0\rangle$.
As input for our analysis, we need the short-distance factors $d_a^n(z,\mu_0)$
and matrix elements $\langle0|{\cal O}_n^H|0\rangle$ for the following
transitions: $g\to J/\psi$, $g\to\chi_{cJ}$, $c\to J/\psi$, and 
$c\to\chi_{cJ}$, which we shall take from the literature.
In the case of the gluon FF's into $J/\psi$ and $\chi_{cJ}$ mesons, we use the
results of Refs.~\cite{bra} and \cite{yua}, respectively.
The corresponding charm-quark FF's may be found in Refs.~\cite{che} and
\cite{tcy}, respectively.

In order to understand our final results, it is necessary to know the relative
importance of the various terms labelled $n$ in Eq.~(\ref{dec}).
On the one hand, this is determined by the magnitude of the short-distance
factors $d_a^n(z,\mu)$ and, on the other hand, by the relative size
of the various matrix elements $\langle0|{\cal O}_n^H|0\rangle$.
The size of the latter may be estimated by how they scale with the relative
velocity $v$ of the charm quarks inside the charmonium state.
The magnitude of the $d_a^n(z,\mu)$ functions is controlled by the lowest
order of $\alpha_s$ that appears in $d_a^n(z,\mu)$.
Thus, to determine the relative importance of the terms in Eq.~(\ref{dec}),
one needs to know the scaling in $v$ of the matrix elements and the order in
$\alpha_s$ of their coefficients.

Let us start with the gluon FF into $J/\psi$, which is a ${}^3\!S_1$ $c\bar c$
state.
Keeping only the leading terms, Eq.~(\ref{dec}) takes the form
\begin{equation}
\label{gpsi}
D_g^{J/\psi}(z,\mu_0)
=\frac{\alpha_s^3(\mu_0)}{m_c^3}d_g^{[\,\underline{1},{}^3\!S_1]}(z)
\langle0|{\cal O}_1^{J/\psi}({}^3\!S_1)|0\rangle
+\frac{\alpha_s(\mu_0)}{m_c^3}d_g^{[\,\underline{8},{}^3\!S_1]}(z)
\langle0|{\cal O}_8^{J/\psi}({}^3\!S_1)|0\rangle,
\end{equation}
where $d_g^{[\,\underline{1},{}^3\!S_1]}(z)$ and
$d_g^{[\,\underline{8},{}^3\!S_1]}(z)$ \cite{bra} are dimensionless functions
listed in the Appendix.
The first term in Eq.~(\ref{gpsi}) corresponds to the colour-singlet
contribution leading in $v$.
Its matrix element $\langle0|{\cal O}_1^{J/\psi}({}^3\!S_1)|0\rangle$ is
proportional to the probability for the formation of a $J/\psi$ meson from a
pointlike $c\bar c$ pair in a color-singlet ${}^3\!S_1$ state.
It is related to the $J/\psi$ radial wave function at the origin,
$R_{J/\psi}(0)$, by \cite{bfy}
\begin{equation}
\langle0|{\cal O}_1^{J/\psi}({}^3\!S_1)|0\rangle=
\frac{9}{2\pi}|R_{J/\psi}(0)|^2,
\end{equation}
up to corrections of relative order $v^4$, and may be extracted from the
partial width of the leptonic decay $J/\psi\to\ell^+\ell^-$.
Following Ref.~\cite{bfy}, we use
$\langle0|{\cal O}_1^{J/\psi}({}^3\!S_1)|0\rangle=1.13$~GeV$^3$, which
includes the $O(\alpha_s)$ correction to the leptonic width.
The leading contribution to the short-distance coefficient
$d_g^{[\,\underline{1},{}^3\!S_1]}(z)$ comes from the parton process
$g^*\to c\bar cg$, where $g^*$ denotes a virtual gluon, and is of order
$\alpha_s^3$.
The matrix element $\langle0|{\cal O}_1^{J/\psi}({}^3\!S_1)|0\rangle$ is of
order $v^3$, so that the contribution to the gluon FF is of order
$\alpha_s^3v^3$.
All other matrix elements in the expansion in Eq.~(\ref{gpsi}) are of higher
order in $v^2$ and would, therefore, be neglected in the colour-singlet
approximation.
However, in the charmonium system, $v^2$ is not actually small; typically one 
has $v^2\approx1/4$, so that not all matrix elements that are formally
suppressed by powers of $v^2$ may be neglected in practice.
Of particular importance is the matrix element
$\langle0|{\cal O}_8^{J/\psi}({}^3\!S_1)|0\rangle$
because it has a short-distance factor of order $\alpha_s$, which
which arises from the parton process $g^*\to c\bar c$.
The corresponding matrix element
$\langle0|{\cal O}_8^{J/\psi}({}^3\!S_1)|0\rangle$ is of order $v^4$.
We take its numerical value to be
$\langle0|{\cal O}_8^{J/\psi}({}^3\!S_1)|0\rangle=0.014$~GeV$^3$ \cite{bfy}.
Thus, the suppression factor $v^4$ is compensated by the factor $1/\alpha_s^2$.
In addition, the different $z$ dependence of
$d_g^{[\,\underline{8},{}^3\!S_1]}(z)$ as compared to the colour-singlet
short-distance coefficient $d_g^{[\,\underline{1},{}^3\!S_1]}(z)$ enhances the
relative importance of the colour-octet contribution.

The gluon FF's into the ${}^3\!P_J$ charmonia $\chi_{cJ}$ have been worked out
in Ref.~\cite{yua} and take the form
\begin{equation}
D_g^{\chi_{cJ}}(z,\mu_0)
=\frac{\alpha_s^2(\mu_0)}{m_c^5}d_g^{[\,\underline{1},{}^3\!P_J]}(z)
\langle0|{\cal O}_1^{\chi_{cJ}}({}^3\!P_J)|0\rangle
+\frac{\alpha_s(\mu_0)}{m_c^3}d_g^{[\,\underline{8},{}^3\!S_1]}(z)
\langle0|{\cal O}_8^{\chi_{cJ}}({}^3\!S_1)|0\rangle.
\end{equation}
The colour-singlet matrix element
$\langle0|{\cal O}_1^{\chi_{cJ}}({}^3\!P_J)|0\rangle$ is related to the
derivative of the nonrelativistic radial wave function of the $P$-wave states
at the origin, $R_{\chi_c}^\prime(0)$, by \cite{bfy}
\begin{equation}
\langle0|{\cal O}_1^{\chi_{cJ}}({}^3\!P_J)|0\rangle
=(2J+1)\frac{9}{2\pi}|R_{\chi_c}^\prime(0)|^2,
\end{equation}
up to corrections of relative order $v^4$.
This parameter can be determined phenomenologically
from the annihilation rates of the $\chi_{cJ}$ mesons.
Since both matrix elements are of the same order in $v$ and the
colour-octet term has one power of $\alpha_s$ less, it is
expected to dominate.
According to Refs.~\cite{yua} and \cite{bfy}, the matrix elements take the
values
\begin{eqnarray}
\label{gchis}
\langle0|{\cal O}_1^{\chi_{c0}}({}^3\!P_0)|0\rangle&\n=\n&\frac{1}{2J+1}
\langle0|{\cal O}_1^{\chi_{cJ}}({}^3\!P_J)|0\rangle=0.0862~\mbox{GeV}^5,\\
\label{gchio}
\langle0|{\cal O}_8^{\chi_{c0}}({}^3\!S_1)|0\rangle&\n=\n&\frac{1}{2J+1}
\langle0|{\cal O}_8^{\chi_{cJ}}({}^3\!S_1)|0\rangle=0.0076~\mbox{GeV}^3.
\end{eqnarray}
In the evaluation of Eq.~(\ref{gchis}), we have used
$2m_c=M_{J/\psi}=3.09688$~GeV \cite{pdg}.
The value of the colour-octet matrix element in Eq.~(\ref{gchio}) is larger
than the most recent value obtained from $B$-meson decays \cite{ebr}.

We now turn to charm-quark fragmentation.
The LO formula for the $c\to J/\psi$ FF has been found in Ref.~\cite{che} and
takes the form
\begin{equation}
\label{cpsi}
D_c^{J/\psi}(z,\mu_0)=\frac{\alpha_s^2(\mu_0)}{m_c^3}
d_c^{[\,\underline{1},{}^3\!S_1]}(z)
\langle0|{\cal O}_1^{J/\psi}({}^3\!S_1)|0\rangle
+\frac{\alpha_s^2(\mu_0)}{m_c^3}
d_c^{[\,\underline{8},{}^3\!S_1]}(z)
\langle0|{\cal O}_8^{J/\psi}({}^3\!S_1)|0\rangle,
\end{equation}
where $d_c^{[\,\underline{1},{}^3\!S_1]}(z)$ and
$d_c^{[\,\underline{8},{}^3\!S_1]}(z)$ may be found in the Appendix.
While, in the case of $g\to J/\psi$, the $v^4$ suppression of
$\langle0|{\cal O}_8^{J/\psi}({}^3\!S_1)|0\rangle$ relative to
$\langle0|{\cal O}_1^{J/\psi}({}^3\!S_1)|0\rangle$ is compensated by
the fact that the colour-octet coefficient is enhanced by a factor
$1/\alpha_s^2$ relative to the colour-singlet one, such a compensation does
not occur in Eq.~(\ref{cpsi}).
In fact, the second term in Eq.~(\ref{cpsi}) only amounts to about 0.12\% of
the first one.

The $c\to\chi_{cJ}$ FF's are given by \cite{tcy}
\begin{equation}
\label{cchi}
D_c^{\chi_{cJ}}(z,\mu_0)
=\frac{\alpha_s^2(\mu_0)}{m_c^5}d_c^{[\,\underline{1},{}^3\!P_J]}(z)
\langle0|{\cal O}_1^{\chi_{cJ}}({}^3\!P_J)|0\rangle
+\frac{\alpha_s^2(\mu_0)}{m_c^3}d_c^{[\,\underline{8},{}^3\!S_1]}(z)
\langle0|{\cal O}_8^{\chi_{cJ}}({}^3\!S_1)|0\rangle,
\end{equation}
where $d_c^{[\,\underline{1},{}^3\!P_J]}(z)$ and
$d_c^{[\,\underline{8},{}^3\!P_J]}(z)$ are specified in the Appendix.
As pointed out in Ref.~\cite{tcy}, the colour-octet term in Eq.~(\ref{cchi})
is minuscule.
Although the minimum invariant mass of the fragmenting charm quark is $3m_c$,
for consistency with the gluon FF's, we choose $\mu_0=2m_c$ in
Eqs.~(\ref{cpsi}) and (\ref{cchi}).

For the calculation of the $J/\psi$ production cross sections, we need the
FF's at the factorization scale $\mu\gg M_{J/\psi}$.
Then, large logarithms in $\mu/M_{J/\psi}$ appear, which have to be resummed.
This is achieved by using the Altarelli-Parisi equations,
\begin{equation}
\label{ape}
\frac{\mu^2d}{d\mu^2}D_a^H(z,\mu)=\sum_b\int_z^1\frac{dx}{x}
P_{ba}^{(T)}\left(\frac{z}{x},\alpha_s(\mu)\right)D_b^H(x,\mu),
\end{equation}
where
\begin{equation}
\label{spl}
P_{ba}^{(T)}\left(z,\alpha_s(\mu)\right)
=\frac{\alpha_s(\mu)}{2\pi}P_{ba}^{(0,T)}(z)
+\left(\frac{\alpha_s(\mu)}{2\pi}\right)^2P_{ba}^{(1,T)}(z)+
O(\alpha_s^3)
\end{equation}
are the timelike splitting functions of parton $a$ into parton $b$.
In our NLO analysis, we include the splitting functions through order 
$\alpha_s^2$ \cite{cur}, while in our LO analysis we truncate Eq.~(\ref{spl})
after the first term.
The initial conditions $D_a^H(z,\mu_0)$ for $a=g,c$ and
$H=J/\psi,\chi_{cJ}$ have been specified above.
Due to charge-conjugation invariance, we have
$D_{\bar c}^H(z,\mu_0)=D_c^H(z,\mu_0)$.
The initial light-quark FF's are set equal to zero.
They are generated at larger scales $\mu$ via Eq.~(\ref{ape}).
However, we expect their effect to be small.
The splitting functions $P_{ba}^{(T)}\left(z,\alpha_s(\mu)\right)$ are for
massless quarks throughout.

In Ref.~\cite{don}, an improved version of Eq.~(\ref{ape}), which respects the
phase-space constraint $D_a^H(z,\mu)=0$ for $z<M_H^2/\mu^2$, has been
suggested.
In Ref.~\cite{lon}, it was found that the numerical effect of this improvement 
is not crucial at large scales.
Since we are mainly interested in the fragmentation production of $J/\psi$
mesons at large $p_T$, for simplicity, we shall stick to Eq.~(\ref{ape}) for
the time being, leaving the implementation of the generalized evolution
equations to future work.

As in our recent work on $D^{*\pm}$ photoproduction \cite{kni}, we start from
the NLO hard-scattering cross sections calculated in the $\overline{\mbox{MS}}$
scheme with massless flavours, and slightly modify the factorization scheme
for the collinear singularities associated with final-state charm quarks.
Specifically, whenever final-state collinear singularities are subtracted from
the hard-scattering cross sections, we substitute
$P_{ca}^{(0,T)}(z)\ln(s/\mu^2)\to P_{ca}^{(0,T)}(z)\ln(s/\mu^2)-d_{ca}(z)$,
where $P_{ca}^{(0,T)}(z)$ are the LO timelike $a\to c$ splitting functions of
Eq.~(\ref{spl}) and the $d_{ca}(z)$ functions may be found in Ref.~\cite{kni}.
Here, $a=g,c,q$, where $q$ stands for the first three quark flavours.
With this procedure, the factorization of the final-state collinear
singularities associated with the charm quark is adjusted so as to match the
finite-$m_c$ calculation.
This is equivalent to the matching approach proposed in Ref.~\cite{mel}
between the massless-charm calculation in connection with the perturbative
FF's and the massive-charm calculation without FF's.

\section{Results}

This section consists of three parts.
Firstly, we shall specify our assumptions concerning the proton and photon
PDF's as well as the equivalent photon approximation.
Secondly, we shall present our predictions for the particular range of
$\gamma p$ CM energies $W$ that is used in the H1 analysis of
inclusive $J/\psi$ production \cite{aid}, namely, 30~GeV${}<W<150$~GeV.
In this subsection, we shall compare the results for the various channels of
interest:
prompt $J/\psi$ photoproduction versus non-prompt $J/\psi$ production
originating from $\chi_{cJ}$ photoproduction with subsequent
$\chi_{cJ}\to J/\psi+\gamma$ decay;
colour-singlet channel versus colour-octet channel;
direct photoproduction versus resolved photoproduction.
In the third part, we consider the high-$W$ range, 150~GeV${}<W<280$~GeV,
where the fragmentation contribution is greatly increased relative to 
the fusion contribution.

\subsection{Input information}

For the calculation of the cross section $d^2\sigma/dy_{lab}\,dp_T$, we adopt
the present HERA conditions, where $E_p = 820$~GeV protons collide with
$E_e = 27.5$~GeV positrons in the laboratory frame, so that
$\sqrt s=2\sqrt{E_pE_e}=300$~GeV is available in the CM frame.
We take the rapidity $y_{lab}$ to be positive in the proton flight direction.
The quasi-real-photon spectrum is described in the Weizs\"acker-Williams
approximation by the formula
\begin{equation}
\label{wwa}
f_{\gamma/e}(x) = \frac{\alpha}{2\pi}\left[\frac{1+(1-x)^2}{x}
\ln\frac{Q_{max}^2}{Q_{min}^2}
+2m_e^2x\left(\frac{1}{Q_{max}^2}-
              \frac{1}{Q_{min}^2}\right)\right],
\end{equation}
where $x=E_\gamma/E_e=W^2/s$, $Q_{min}^2=m_e^2x^2/(1-x)$,
$\alpha$ is Sommerfeld's fine-structure constant, and $m_e$ is the electron
mass.
We consider the case where the final-state electron is not tagged, so that
$Q_{max}^2=4$~GeV$^2$.
The $\gamma p$ energy interval 30~GeV${}<W<150$~GeV (150~GeV${}<W<280$~GeV)
considered in Section~3.2 (3.3) corresponds to $0.010<x<0.249$
($0.249<x<0.869$).

We work at NLO in the $\overline{\mbox{MS}}$ scheme with $n_f=4$ flavours.
As for the proton and photon PDF's, we use set CTEQ4M \cite{lai}, with
$\Lambda_{\overline{\mbox{\scriptsize MS}}}^{(4)}=296$~MeV, and set GRV~HO
\cite{grv} after converting it from the DIS${}_\gamma$ scheme to the
$\overline{\mbox{MS}}$ scheme.
We identify the factorization scales associated with the proton, photon and
final-state hadron and collectively denote them by $M_f$.
For $M_f$ and the renormalization scale $\mu$, we choose $\mu=M_f=m_T$,
where $m_T$ is the $J/\psi$ transverse mass defined above.
We take the starting scale of the FF's to be $\mu_0=2m_c$.
We calculate $\alpha_s(\mu)$ from the two-loop formula with
$\Lambda_{\overline{\mbox{\scriptsize MS}}}^{(4)}$ equal to the value used in
the proton PDF's.
Unfortunately, the NLO correction to the perturbative coefficient functions
discussed above, which, strictly speaking, ought to be included as well, are
not yet available.
However, in analogy to the genuine NLO corrections to $J/\psi$
photoproduction at HERA in the CSM, which are only at the 20\% level, these 
are likely to be moderate as well.

The LO results which enter the QCD-correction ($K$) factors are calculated 
consistently, {\it i.e.}, using the one-loop formula for $\alpha_s(\mu)$,
the CTEQ4L \cite{lai} proton PDF's, with
$\Lambda_{\overline{\mbox{\scriptsize MS}}}^{(4)}=236$~MeV, and
the GRV~LO photon PDF's, and evolving the FF's with the LO splitting functions.
In all calculations, we employ the numerical values of the matrix elements
specified above.

\subsection{Results for the low-$W$ range}

In this subsection, we investigate the NLO cross section
$d^2\sigma/dy_{lab}\,dp_T$ of inclusive $J/\psi$ photoproduction at HERA,
integrated over 30~GeV${}<W<150$~GeV as in the H1 experiment \cite{aid}.
In order to analyze the $p_T$ dependence of the cross section in the central
region of the detector, we integrate $d^2\sigma/dy_{lab}\,dp_T$ over
$-1.5<y_{lab}<1$.
The result is displayed in Figs.~\ref{fig1}(a)--(c) for direct and resolved 
photoproduction and their sum, respectively.
In each case, we show how the total fragmentation contributions (solid lines)
are composed of the respective prompt colour-singlet (dashed lines), prompt
colour-octet (dot-dashed lines), and non-prompt (dot-dot-dashed lines)
contributions.
For comparison, we also show the LO direct-photon contribution of the CSM
(dotted lines).
We have to bear in mind that the latter is enhanced by typically 20\% due to
genuine NLO QCD corrections to the parton-fusion cross section \cite{kra}.
However, we implicitly include the bulk of the QCD corrections, namely, those
to the leptonic $J/\psi$ decay width, which affect the value of
$\langle0|{\cal O}_1^{J/\psi}({}^3\!S_1)|0\rangle$ extracted from experiment
\cite{bfy}.
For consistency with the fragmentation contribution, we also choose
$\mu=M_f=m_T$ in the fusion contribution.
We do not take into account the resolved-photon CSM contribution, which
appreciably contributes only at very low $\tilde z$, {\it i.e.}, at low $p_T$
and large $y_{lab}$ \cite{jun}.
A dedicated study of the low-$\tilde z$ range, which we intend to do in the 
future, should also consider this contribution.

As anticipated in the Introduction, both in the direct- and resolved-photon
channels, the fragmentation cross sections fall off less steeply with $p_T$
than the fusion cross section.
In all cases, we observe that the prompt colour-octet contribution is most
significant and dominates the large-$p_T$ behaviour.
The non-prompt cross section is always smaller than the prompt cross section.
In the sum of the direct- and resolved-photon channels, their ratio ranges
from 2.8 at $p_T=3$~GeV to 4.3 at $p_T=15$~GeV.
We emphasize that, in the low-$p_T$ range, the resolved-photon component,
which was previously neglected \cite{god}, is considerably more important than
the direct-photon component, by a factor of 18 (2.3) at $p_T=3$~GeV (10~GeV).
The two components cross over at $p_T\approx15$~GeV.
Notice also that, even for $p_T\simlt6$~GeV, the resolved-photon component is
comparable to the fusion contribution;
for $p_T>7$~GeV, it exceeds the fusion contribution.
In the resolved-photon contribution, the prompt colour-singlet part is
negligibly small compared to the prompt colour-octet part.
This is caused by the dominance of $g\to J/\psi+X$ in the resolved-photon cross
section and the suppression of colour-singlet gluon fragmentation mentioned
above.
In the direct-photon cross section, this hierarchy is reduced since
$c\to J/\psi+X$ is more important.

In general, rapidity distributions allow for a more specific investigation of
the individual contributions than transverse-momentum distributions.
In Figs.~\ref{fig2}(a)--(c), the $y_{lab}$ dependence of
$d^2\sigma/dy_{lab}\,dp_T$ is displayed at $p_T=5$ and 10~GeV for the same
cases as in Figs.~\ref{fig1}(a)--(c).
While the direct-photon component peaks in the same $y_{lab}$ region as the
fusion contribution, the resolved-photon cross section dominantly contributes
in the forward direction, with a maximum at about $y_{lab}=2.5$.
As might be inferred from Figs.~\ref{fig1}(a)--(c), the prompt colour-octet
contribution is dominant; in fact, this is true over the full $y_{lab}$ range.
We observe that the $y_{lab}$ spectra of the prompt colour-singlet channel,
although suppressed, have a shape similar to the fusion contribution and are
peaked at significantly smaller $y_{lab}$ values than the other fragmentation
channels.
At large $p_T$, the dominant $J/\psi$ production mechanism in $p\bar p$
collisions at the Tevatron is colour-octet gluon fragmentation \cite{roy}.
In order for HERA experiments to probe this component, it will be necessary
to take data in the very forward direction, where the resolved-photon
colour-octet contribution is most significant.
It should be clear that the colour-octet contributions shown in
Figs.~\ref{fig2}(a)--(c) are proportional to the corresponding matrix elements
specified in Section~2, which are extracted from fits to Tevatron data.
So, in order to substantiate (or question) the huge prompt colour-octet
$J/\psi$ contribution that is apparently indispensable to reconcile the
Tevatron data with theory, it would be desirable to collect data in the very
forward region at HERA, which has not yet been explored by the H1 and ZEUS
collaborations.

As mentioned in the Introduction, the present analysis extends the previous
work \cite{god} on $J/\psi$ fragmentation production at HERA in two important
respects.
For one thing, we include the resolved-photon contribution, which was not 
mentioned in Ref.~\cite{god}.
On the other hand, we work at NLO, while the analysis of Ref.~\cite{god} 
proceeds at LO.
Above, we have stressed the significance of the resolved-photon contribution,
which is nicely illustrated in Figs.~\ref{fig1}(b) and \ref{fig2}(b).
In the following, we shall assess the impact of the NLO corrections on the
various fragmentation contributions.
To this end, we consistently repeat the analysis of Figs.~\ref{fig1}(a)--(c)
at LO, as described in Section~2, and plot the resulting NLO to LO ratios in
Figs.~\ref{fig3}(a)--(c).
In the case of direct photoproduction considered in Fig.~\ref{fig3}(a), the
prompt colour-singlet contribution has an approximately constant $K$ factor of
about 0.6, while the prompt colour-octet and non-prompt contributions have
diminishing $K$ factors only at $p_T$ values below 6.5 and 4, respectively.
In the case of resolved photoproduction addressed in Fig.~\ref{fig3}(b), the
$K$ factors of the various channels are always significantly larger than 
unity, and the overall $K$ factor ranges between 2 and 2.4.
In both the direct- and resolved-photon cases, the $K$ factors of the
non-prompt contributions take extraordinarily large values in the upper $p_T$
range.
Detailed investigation reveals that this enhancement essentially originates
from the lower edge of the kinematically allowed $y_{lab}$ range.
However, since, at large $p_T$, these contributions are greatly suppressed
relative to the prompt colour-octet contributions, this is inconsequential for
the superposition of all channels, which is experimentally observable.
In fact, the overall $K$ factor, which is shown as the solid line in 
Fig.~\ref{fig3}(c), is well behaved, ranging between 1.6 and 1.8 in the $p_T$
range considered.

At this point, we should comment on the LO analysis of direct $J/\psi$
photoproduction via fragmentation at HERA reported in Ref.~\cite{god}.
Similarly to the present study, the authors of Ref.~\cite{god} included the
prompt, non-prompt, colour-singlet, and colour-octet contributions.
However, they did not convolute the $\gamma p$ cross section with the
Weizs\"acker-Williams distribution, but considered fixed photon energies
instead.
There is an overall factor of 18 missing on the right-hand side of Eq.~(3) in
Ref.~\cite{god}, which is to describe the fusion cross section.
In the case of fragmentation production, a meaningful quantitative comparison
is impossible, since the input values of the poorly known colour-octet matrix
elements are not to be found in Ref.~\cite{god}.

One should bear in mind that all results presented in this subsection refer to
the low-$W$ range 30~GeV${}<W<150$~GeV.
We expect that the resolved-photon contribution will be enhanced relative to
the fusion contribution if $W$ is increased.
In the next subsection, this issue will be addressed in some detail.

\subsection{Results for the high-$W$ range}

In order to test the predictions for $J/\psi$ photoproduction via 
fragmentation at HERA that follow from the theoretical analysis of the
Tevatron data, it is useful to find regions of phase space where the fusion
mechanism is suppressed.
As we have seen in Figs.~\ref{fig2}(a)--(c), this happens, for instance, in
the very forward region of the detector.
But even in the central region, $-1.5<y_{lab}<1$, to which previous 
measurements of charmed-meson photoproduction at HERA were confined
\cite{kar}, fragmentation production may be enhanced with respect to the fusion
mechanism by selecting events with high photon energy.
This is demonstrated in Fig.~\ref{fig4}, where the $W$ dependence of
$d^2\sigma/dW\,dp_T$ is shown for $p_T=5$ and 10~GeV.
While the $W$ distribution of the direct-photon contribution has a shape
similar to that of the fusion contribution, with a maximum at low to
intermediate $W$, the resolved-photon contribution tends to be monotonically
increasing with increasing $W$.
As a consequence, even at low $p_T$, fragmentation production becomes more 
important than the fusion mechanism beyond some point in $W$.
At $p_T=5$, the cross-over takes place at about $W=125$~GeV, which corresponds 
to $x=0.17$ in Eq.~(\ref{wwa}).

Guided by this observation, we repeat the analysis of Fig.~\ref{fig1}(c)
integrating over 150~GeV ${}<W<280$~GeV, instead of 30~GeV${}<W<150$~GeV, and
plot the outcome in Fig.~\ref{fig5}.
As expected, this leads to a significant increase of the fragmentation 
contribution, which mainly originates from the resolved-photon part.
On the other hand, the fusion contribution is only moderately increased at
$p_T\simgt4$~GeV, while it is even decreased at lower values of $p_T$.
In conclusion, over the entire $p_T$ range, the fragmentation contribution is
now more than twice as large as the fusion contribution;
at $p_T\simgt11$~GeV, the fusion contribution falls short of the fragmentation
contribution by more than one order of magnitude.

At HERA, it should also be possible to detect $J/\psi$ mesons in the very 
forward direction, almost up to $y_{lab}=3.5$ \cite{kie}.
This would offer the opportunity to render the fragmentation to fusion ratio 
even more favourable.
Thus, in Fig.~\ref{fig6}, we investigate how the results for
150~GeV${}<W<280$~GeV in Fig.~\ref{fig5} are modified if the $y_{lab}$
interval is shifted from $-1.5<y_{lab}<1$ to $1<y_{lab}<3.5$.
We observe that the fragmentation contribution is then enhanced, almost by a
factor of two at large $p_T$.
But, what is even more significant, the fusion contribution is reduced by one
to two orders of magnitude.
As a consequence, the fragmentation to fusion ratio is dramatically increased,
to values in excess of 200 throughout the entire $p_T$ range.
We conclude, that the kinematical region of large $W$ and large $y_{lab}$ is
an ideal place to probe at HERA the anomalously large inclusive $J/\psi$ yield
observed at the Tevatron.

It has become customary to present experimental data on inclusive $J/\psi$
production in terms of the inelasticity variable
$\tilde z$ defined in the Introduction, instead of using $y_{lab}$.
The relation between $\tilde z$ and $y_{lab}$ reads
\begin{equation}
\tilde z=\frac{2E_pm_T}{W^2}e^{-y_{lab}},
\end{equation}
where $E_p$ is the proton energy in the laboratory system.
Thus, the kinematical range of interest corresponds to small values of
$\tilde z$.
This is in line with Ref.~\cite{god}, where the kinematical cut $\tilde z<0.5$
has been proposed to enhance the fragmentation contribution relative to the
fusion contribution.
We have recently assessed, in a NLO analysis \cite{let}, the logistics of
enhancing via appropriate acceptance cuts the colour-octet contribution to
$J/\psi$ production via fragmentation at HERA.

\section{Conclusions}

In addition to the well-known parton-fusion mechanism, $J/\psi$ mesons may 
also be produced via the fragmentation of final-state partons.
The latter mechanism is expected to be the dominant source of $J/\psi$ mesons
at sufficiently large $p_T$.
In this paper, we studied the photoproduction of $J/\psi$ mesons at large
$p_T$ via fragmentation at HERA.

Our calculation was performed at NLO in the QCD-improved parton model with
$n_f=4$ massless quark flavours, {\it i.e.}, the charm quark was assumed to be 
an active flavour inside the proton and the resolved photon.
We employed charm and gluon FF's which were perturbatively calculated at a
fragmentation scale of order $M_{J/\psi}$ in nonrelativistic QCD, and evolved 
them in $z$ space to the characteristic scale of the considered process, of
order $p_T$, using NLO Altarelli-Parisi splitting functions.
At the same time, we adjusted the factorization scheme of the collinear
singularities connected with final-state charm quarks so as to match the
corresponding calculation with massive charm quarks.
In this way, large logarithms of the type $\alpha_s\ln(p_T^2/M_{J/\psi}^2)$,
which would be present in the massive calculation and render it unreliable at
large $p_T$, are removed from the hard-scattering cross sections and properly
resummed.
We included prompt $J/\psi$ fragmentation production as well as $\chi_{cJ}$
fragmentation production followed by $\chi_{cJ}\to J/\psi+\gamma$ decays
(non-prompt $J/\psi$ fragmentation production).
In each channel, we took into account the contributions due to
colour-singlet and colour-octet $c\bar c$ states.
Apart from direct photoproduction, which had been considered to LO in 
Ref.~\cite{god}, we also studied resolved photoproduction, both at NLO.

While the nonperturbative matrix elements multiplying the colour-singlet
FF's may be reliably extracted from the decay properties of the charmonium
states, the colour-octet matrix elements are adjusted to account for the 
enormous excess of recent Tevatron data on large-$p_T$ charmonium production
over the theoretical prediction within the CSM.
This procedure implicitly assumes, on the theoretical side, that there are no
other production mechanisms that have been overlooked so far and, on the 
experimental side, that all backgrounds are fully under control.
Therefore, the colour-octet matrix elements must be considered far less
rigorously determined than the colour-singlet matrix elements.
Obviously, it is of prime importance to probe the colour-octet matrix elements
in other experiments as well.
The signature for colour-octet $J/\psi$ production in $e^+e^-$ annihilation at 
the $Z$-boson resonance has recently been investigated in Ref.~\cite{lon,keu}.
Here, we continued this research programme by assessing the prospects of the
HERA experiments to substantiate the findings at the Tevatron.

Our main observations may be summarized as follows.
As for direct $J/\psi$ photoproduction, fragmentation production starts to 
have a higher yield than the fusion mechanism of the CSM at 
$p_T\simgt10$~GeV [see Figs.~\ref{fig1}(a) and \ref{fig2}(a)].
This is in line with the main conclusion of the LO analysis of Ref.~\cite{god},
since the NLO corrections are relatively modest in this case
[see Fig.~\ref{fig3}(a)].
However, in the range $p_T\simlt10$~GeV, which will be accessed at HERA in the 
near future, resolved $J/\psi$ photoproduction via fragmentation, which has 
not yet been considered in the literature, is considerably more significant
than direct photoproduction [see Figs.~\ref{fig1}(b) and \ref{fig2}(b)],
by a factor of 18 (2.3) at $p_T=3$~GeV (10~GeV).
In fact, the resolved-photon fragmentation cross section already exceeds the
fusion cross section at $p_T\simgt5$~GeV and is comparable to it at smaller
values of $p_T$.
The bulk of this cross section originates in the subprocesses
$g_\gamma+q_p\to g+X$ and $q_\gamma+g_p\to g+X$ followed by the fragmentation
chain $g\to c\bar c[\,\underline{8},{}^3\!S_1]\to J/\psi$.
The dominance of resolved photoproduction is partly due to a large QCD $K$
factor, in excess of 2 [see Fig.~\ref{fig3}(b)].
These results refer to the standard ranges $-1.5<y_{lab}<1$ and
30~GeV$<W<150$~GeV.

We suggested two ways to further enhance the (resolved-photon) fragmentation
production relative to the fusion mechanism.
One is to concentrate on the forward direction, {\it e.g.}, $1<y_{lab}<3.5$,
where the resolved-photoproduction cross section peaks [see 
Fig.~\ref{fig2}(b)].
The other one is to increase the CM energy available for the hard $\gamma p$
scattering, {\it e.g.}, by choosing 150~GeV$<W<280$~GeV, as the
resolved-photoproduction cross section increases monotonically with $W$, while
the fusion cross section takes its maximum at relatively low $W$ values
(see Fig.~\ref{fig4}).
The usefulness of increasing $W$ and $y_{lab}$ is nicely illustrated in
Figs.~\ref{fig5} and \ref{fig6}, respectively.
Both options should be feasible at HERA, and we look forward to the exciting
scrutiny of the Tevatron $J/\psi$ colour-octet phenomenon by the HERA 
experiments in the near future.

\bigskip
\centerline{\bf ACKNOWLEDGMENTS}
\smallskip\noindent
We are grateful to Reinhold R\"uckl for raising our interest in this study and
for beneficial discussions at its early stages, to Eric Braaten for providing
a computer code for the evaluation of $d_g^{[\,\underline{1},{}^3\!S_1]}(z)$
\cite{bra}, to Eric Braaten and Sean Fleming for helpful technical advice
concerning the numerical evolution of singular distributions in $z$ space, and
to Kingman Cheung for a useful comment on the suppression of the colour-octet
contribution to the $c\to J/\psi$ FF.
One of us (G.K.) is grateful to the Theory Group of the
Werner-Heisenberg-Institut for the hospitality extended to him during a visit
when this paper was prepared.

\begin{appendix}

\section{Appendix: Coefficients of the perturbative fragmentation functions}

For the reader's convenience, we collect here the colour-singlet and
colour-octet coefficient functions, at the starting scale $\mu_0$, for the
gluon and charm-quark FF's into $J/\psi$ and $\chi_{cJ}$ mesons, with
$J=0,1,2$.
The $g\to c\bar c[\,\underline{1},{}^3\!S_1]$ coefficient function has the
two-dimensional integral representation \cite{bra}
\begin{eqnarray}
d_g^{[\,\underline{1},{}^3\!S_1]}(z)
&\n=\n& {5 \over 5184 \pi}
\int_0^z dr \int_{(r+z^2)/(2z)}^{(1+r)/2} dy
{1 \over (1-y)^2 (y-r)^2 (y^2-r)^2}
\nonumber \\
&\n\n&{}\times \sum_{i=0}^2 z^i \left( f_i(r,y) + g_i(r,y)
	{1+r-2y \over 2 (y-r) \sqrt{y^2-r}}
	\ln{y-r + \sqrt{y^2-r} \over y-r - \sqrt{y^2-r}} \right),
\end{eqnarray}
where
\begin{eqnarray}
f_0(r,y) &\n=\n& r^2(1+r)(3+12r+13r^2) - 16r^2(1+r)(1+3r)y
\nonumber \\
&\n\n&{}- 2r(3-9r-21r^2+7r^3)y^2
+ 8r(4+3r+3r^2)y^3 - 4r(9-3r-4r^2)y^4
\nonumber \\
&\n\n&{}-  16(1+3r+3r^2)y^5 + 8(6+7r)y^6 - 32 y^7 ,
\nonumber \\
f_1(r,y) &\n=\n& -2r(1+5r+19r^2+7r^3)y + 96r^2(1+r)y^2
+ 8(1-5r-22r^2-2r^3)y^3
\nonumber \\
&\n\n&{}+ 16r(7+3r)y^4 - 8(5+7r)y^5 + 32y^6 ,
\nonumber \\
f_2(r,y) &\n=\n& r(1+5r+19r^2+7r^3) - 48r^2(1+r)y - 4(1-5r-22r^2-2r^3)y^2
\nonumber \\
&\n\n&{}- 8r(7+3r)y^3 + 4(5+7r)y^4 - 16y^5 ,
\nonumber \\
g_0(r,y) &\n=\n& r^3(1-r)(3+24r+13r^2) - 4r^3(7-3r-12r^2)y
- 2r^3(17+22r-7r^2)y^2
\nonumber \\
&\n\n&{}+ 4r^2(13+5r-6r^2)y^3 - 8r(1+2r+5r^2+2r^3)y^4
- 8r(3-11r-6r^2)y^5
\nonumber \\
&\n\n&{}+ 8(1-2r-5r^2)y^6 ,
\nonumber \\
g_1(r,y) &\n=\n& -2r^2(1+r)(1-r)(1+7r)y + 8r^2(1+3r)(1-4r)y^2
\nonumber \\
&\n\n&{}+ 4r(1+10r+57r^2+4r^3)y^3
- 8r(1+29r+6r^2)y^4 - 8(1-8r-5r^2)y^5 ,
\nonumber \\
g_2(r,y) &\n=\n& r^2(1+r)(1-r)(1+7r) - 4r^2(1+3r)(1-4r)y
\nonumber \\
&\n\n&{}- 2r(1+10r+57r^2+4r^3)y^2
+ 4r(1+29r+6r^2)y^3 + 4(1-8r-5r^2)y^4 .\qquad
\end{eqnarray}
All other coefficient functions are available in closed form
\cite{bra,yua,che,tcy}:
\begin{eqnarray}
d_g^{[\,\underline{8},{}^3\!S_1]}(z)
&\n=\n&{\pi\over24}\delta(1-z),
\nonumber\\
d_g^{[\,\underline{1},{}^3\!P_0]}(z)
&\n=\n&\frac{2}{81}\left[\frac{1}{4}\delta(1-z)+\left(\frac{1}{1-z}\right)_+
-1+\frac{85}{8}z-\frac{13}{4}z^2+\frac{9}{4}(5-3z)\ln(1-z)\right],
\nonumber\\
d_g^{[\,\underline{1},{}^3\!P_1]}(z)
&\n=\n&\frac{2}{81}\left[\frac{1}{8}\delta(1-z)+\left(\frac{1}{1-z}\right)_+
-1-\frac{1}{4}z-z^2\right],
\nonumber\\
d_g^{[\,\underline{1},{}^3\!P_2]}(z)
&\n=\n&\frac{2}{81}\left[\frac{7}{40}\delta(1-z)+\left(\frac{1}{1-z}\right)_+
-1+\frac{11}{4}z-z^2+\frac{9}{5}(2-z)\ln(1-z)\right],
\nonumber\\
d_c^{[\,\underline{1},{}^3\!S_1]}(z)
&\n=\n&\frac{32}{3}d_c^{[\,\underline{8},{}^3\!S_1]}(z)
=\frac{16z(1-z)^2}{243(2-z)^6}\left(16-32z+72z^2-32z^3+5z^4\right),
\nonumber\\
d_c^{[\,\underline{1},{}^3\!P_0]}(z)
&\n=\n&\frac{16z(1-z)^2}{729(2-z)^8}
\left( 192 + 384z + 528z^2 - 1376z^3 + 1060z^4 - 376z^5 + 59z^6 \right),
\nonumber\\
d_c^{[\,\underline{1},{}^3\!P_1]}(z)
&\n=\n&\frac{64z(1-z)^2}{729(2-z)^8}
\left( 96 - 288z + 496z^2 - 408z^3 + 202z^4 - 54z^5 + 7z^6 \right),
\nonumber\\
d_c^{[\,\underline{1},{}^3\!P_2]}(z)
&\n=\n&\frac{128z(1-z)^2}{3645(2-z)^8}
\left( 48 - 192z + 480z^2 - 668z^3 + 541z^4 - 184z^5 + 23z^6 \right).
\end{eqnarray}
As usual, the plus distribution $1/(1-z)_+$ is defined by
$\int_0^1dz\,f(z)/(1-z)_+=\int_0^1dz\,[f(z)-f(1)]/(1-z)$ for any
regular function $f(z)$.

For the numerical solution of the $\mu^2$-evolution equations in $z$ space, it
is useful to approximate
\begin{eqnarray}
\label{app}
\delta(1-z)&\n=\n&
\cases{
\displaystyle
0,
&if $z\le1-\epsilon$,\cr
&\cr
\displaystyle
\frac{1}{\epsilon},
&if $1-\epsilon<z\le1$,\cr}
\nonumber\\ \nonumber\\
\left(\frac{1}{1-z}\right)_+&\n=\n&
\cases{
\displaystyle
\frac{1}{1-z},
&if $z\le1-\epsilon$,\cr
&\cr
\displaystyle
\frac{\ln\epsilon}{\epsilon},
&if $1-\epsilon<z\le1$,\cr}
\nonumber\\ \nonumber\\
\ln(1-z)&\n=\n&
\cases{
\displaystyle
\ln(1-z),
&if $z\le1-\epsilon$,\cr
&\cr
\displaystyle
\ln\epsilon-1,
&if $1-\epsilon<z\le1$,\cr}
\end{eqnarray}
with appropriate $0<\epsilon\ll1$.
We have verified, for the perturbative FF's of Ref.~\cite{mel} in LO and NLO,
that the $\mu^2$ evolution in $z$ space implemented using Eq.~(\ref{app})
agrees very well with the $\mu^2$ evolution in Mellin space \cite{spi}.

\end{appendix}

\newpage

\vskip-6cm

\begin{figure}

\centerline{\bf FIGURE CAPTIONS}

\caption{\protect\label{fig1} (a) Direct-photon, (b) resolved-photon, and (c)
total contributions to the cross section $d\sigma/dp_T$ of inclusive $J/\psi$
photoproduction via fragmentation at HERA, integrated over
30~GeV${}<W<150$~GeV and $-1.5<y_{lab}<1$.
The NLO contributions due to prompt singlet (P1), prompt octet (P8),
non-prompt singlet plus octet fragmentation (N1+8), and their sum (frag.) are
compared with the LO CSM contribution (fusion).\hskip9cm}

\vskip-.2cm

\caption{\protect\label{fig2} (a) Direct-photon, (b) resolved-photon, and (c)
total contributions to the cross section $d\sigma/dy_{lab}\,p_T$ of inclusive
$J/\psi$ photoproduction via fragmentation at HERA, integrated over
30~GeV${}<W<150$~GeV, at $p_T=5$ and 10~GeV.
The NLO contributions due to prompt singlet (P1), prompt octet (P8),
non-prompt singlet plus octet fragmentation (N1+8), and their sum (frag.) are
compared with the LO CSM contribution (fusion).\hskip9cm}

\vskip-.2cm

\caption{\protect\label{fig3} NLO to LO ratios for the (a) direct-photon, (b)
resolved-photon, and (c) total contributions to the cross section
$d\sigma/dp_T$ of inclusive $J/\psi$ photoproduction via fragmentation at
HERA, integrated over 30~GeV${}<W<150$~GeV and $-1.5<y_{lab}<1$.
The ratios are taken for prompt singlet (P1), prompt octet (P8), non-prompt
singlet plus octet fragmentation (N1+8), and their sum (frag.).\hskip9cm}

\vskip-.2cm

\caption{\protect\label{fig4} Cross section $d^2\sigma/dW\,dp_T$ of inclusive
$J/\psi$ photoproduction via fragmentation at HERA, integrated over
$-1.5<y_{lab}<1$, at $p_T=5$ and 10~GeV.
The NLO fragmentation contributions due direct photons, resolved photons, and
their sum are compared with the LO CSM contribution.\hskip9cm}

\vskip-.2cm

\caption{\protect\label{fig5} NLO fragmentation and LO CSM
contributions to the cross section $d\sigma/dp_T$ of inclusive $J/\psi$
photoproduction at HERA, integrated over $-1.5<y_{lab}<1$ and two different
$W$ intervals, namely, 30~GeV${}<W<150$~GeV and 150~GeV${}<W<280$~GeV.
\hskip9cm}

\vskip-.2cm

\caption{\protect\label{fig6} NLO fragmentation and LO CSM
contributions to the cross section $d\sigma/dp_T$ of inclusive $J/\psi$
photoproduction at HERA, integrated over 150~GeV${}<W<280$~GeV and two
different $y_{lab}$ intervals, namely, $-1.5<y_{lab}<1$ and $1<y_{lab}<3.5$.
\hskip9cm}

\end{figure}

\newpage
\begin{figure}[ht]
\epsfig{figure=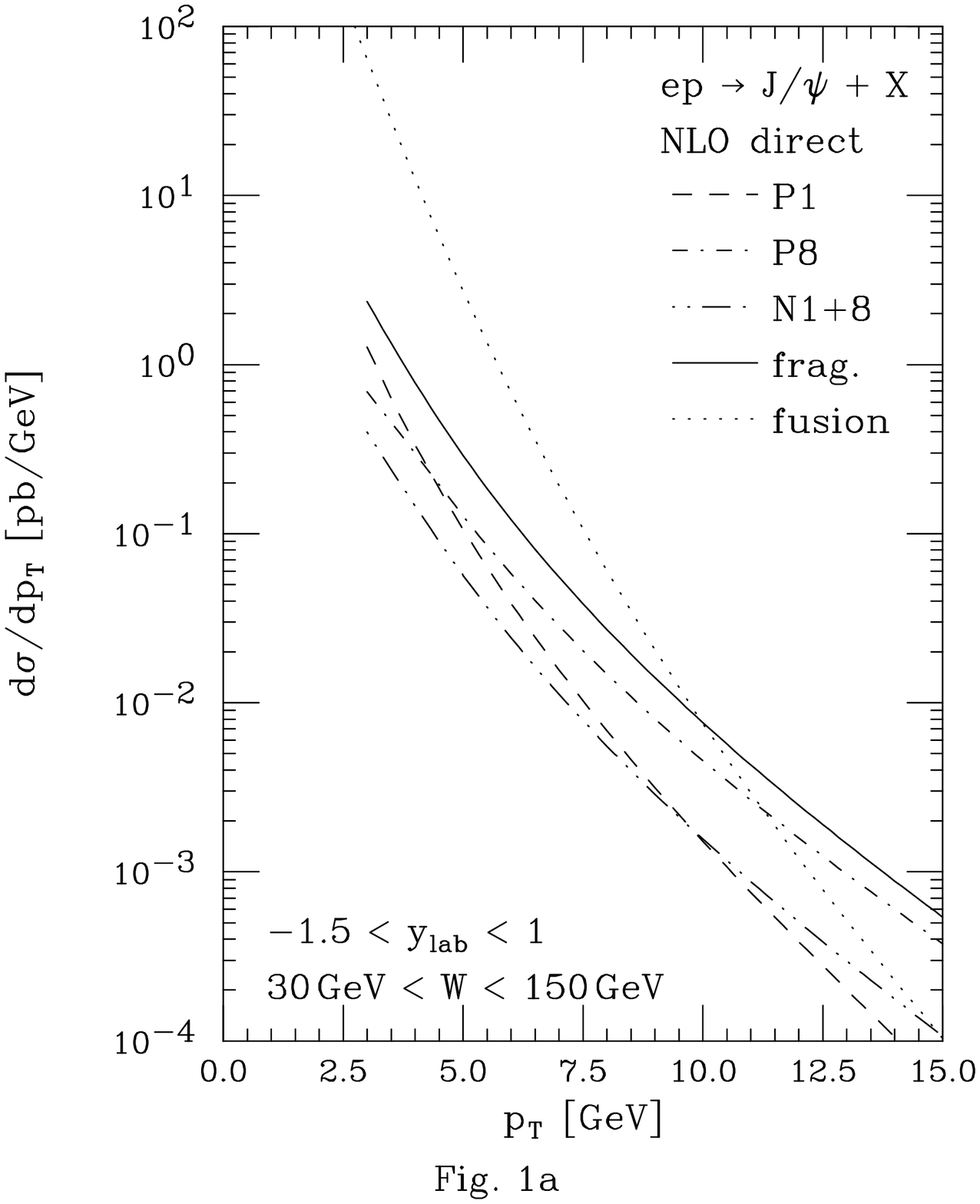,width=\textwidth}
\end{figure}
\newpage
\begin{figure}[ht]
\epsfig{figure=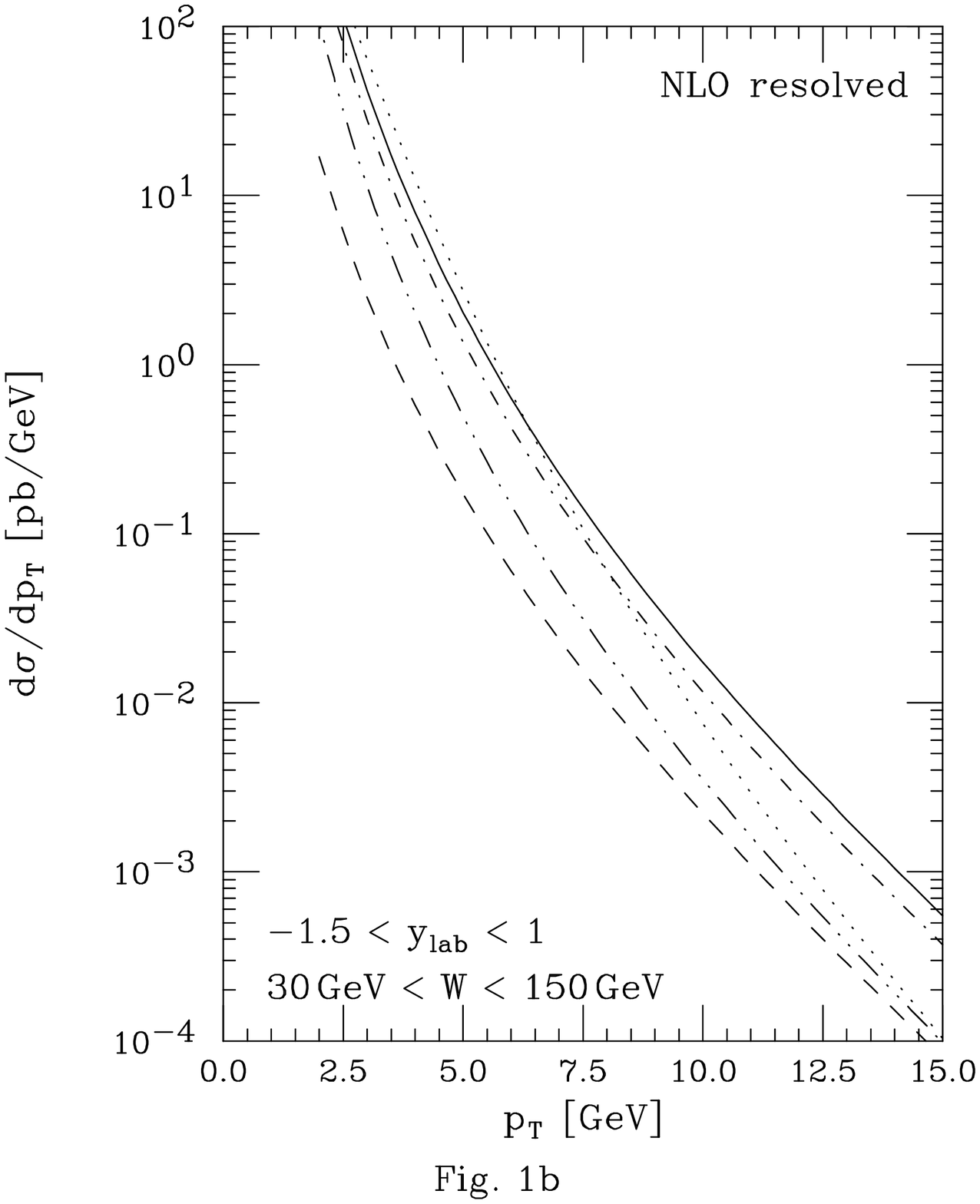,width=\textwidth}
\end{figure}
\newpage
\begin{figure}[ht]
\epsfig{figure=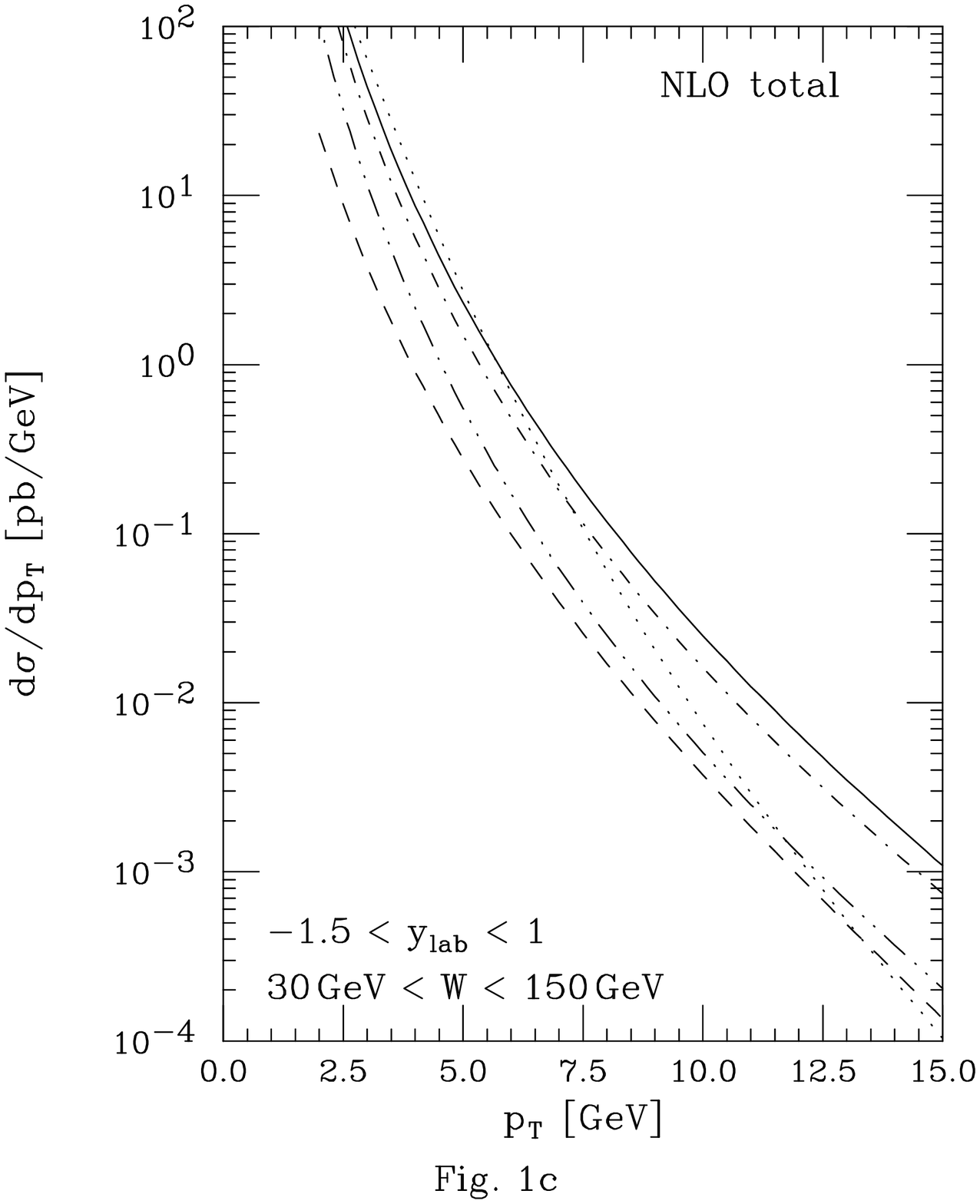,width=\textwidth}
\end{figure}
\newpage
\begin{figure}[ht]
\epsfig{figure=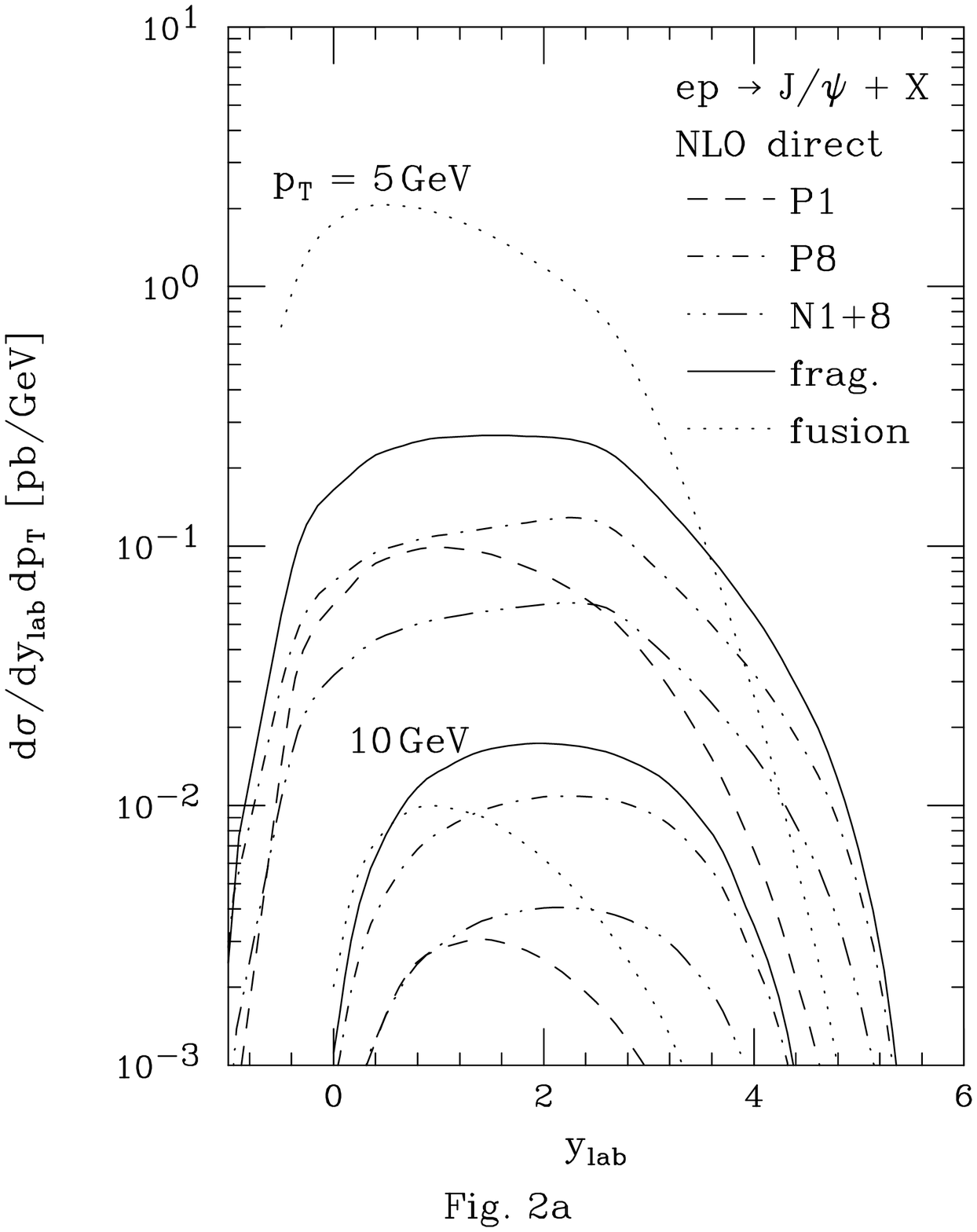,width=\textwidth}
\end{figure}
\newpage
\begin{figure}[ht]
\epsfig{figure=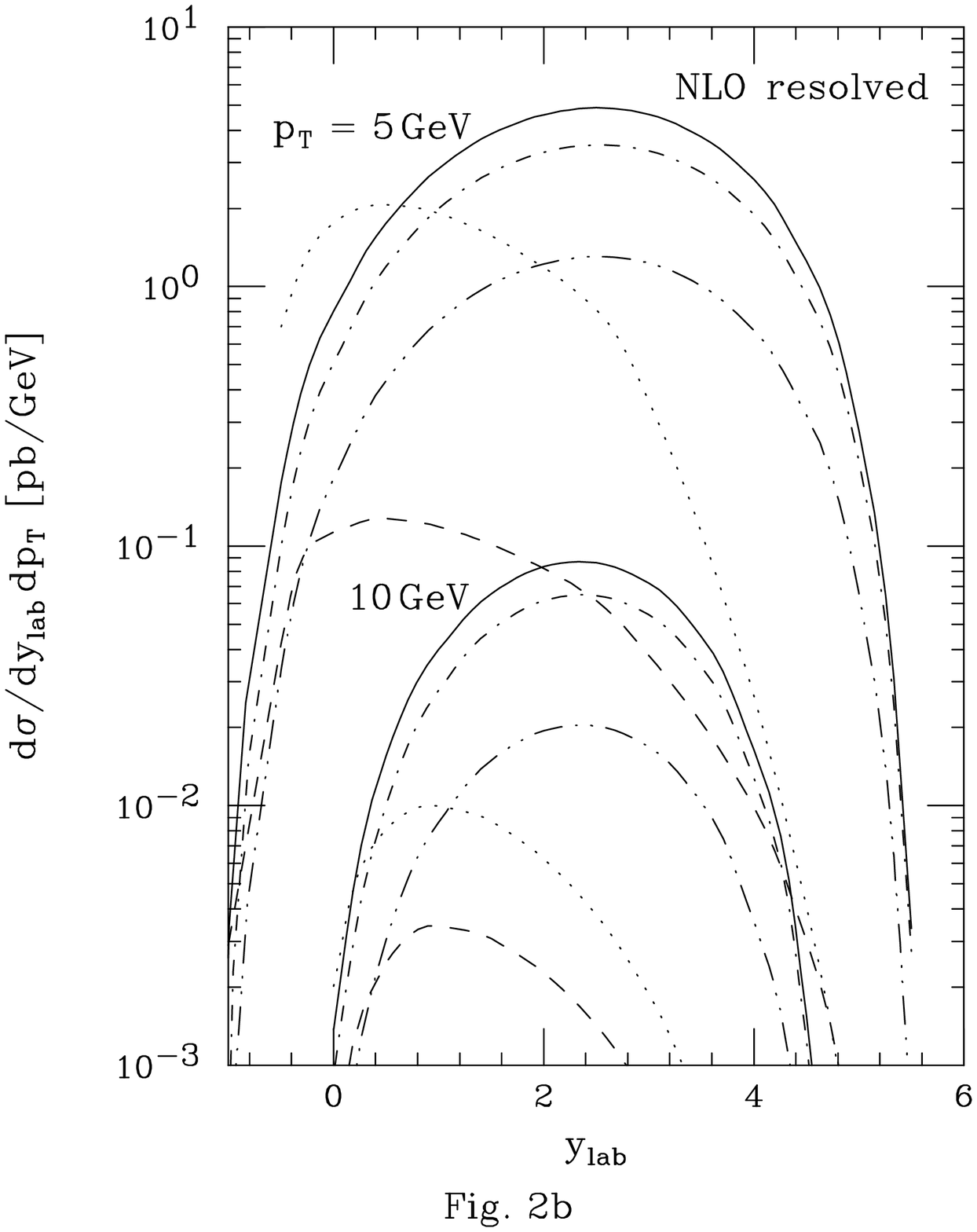,width=\textwidth}
\end{figure}
\newpage
\begin{figure}[ht]
\epsfig{figure=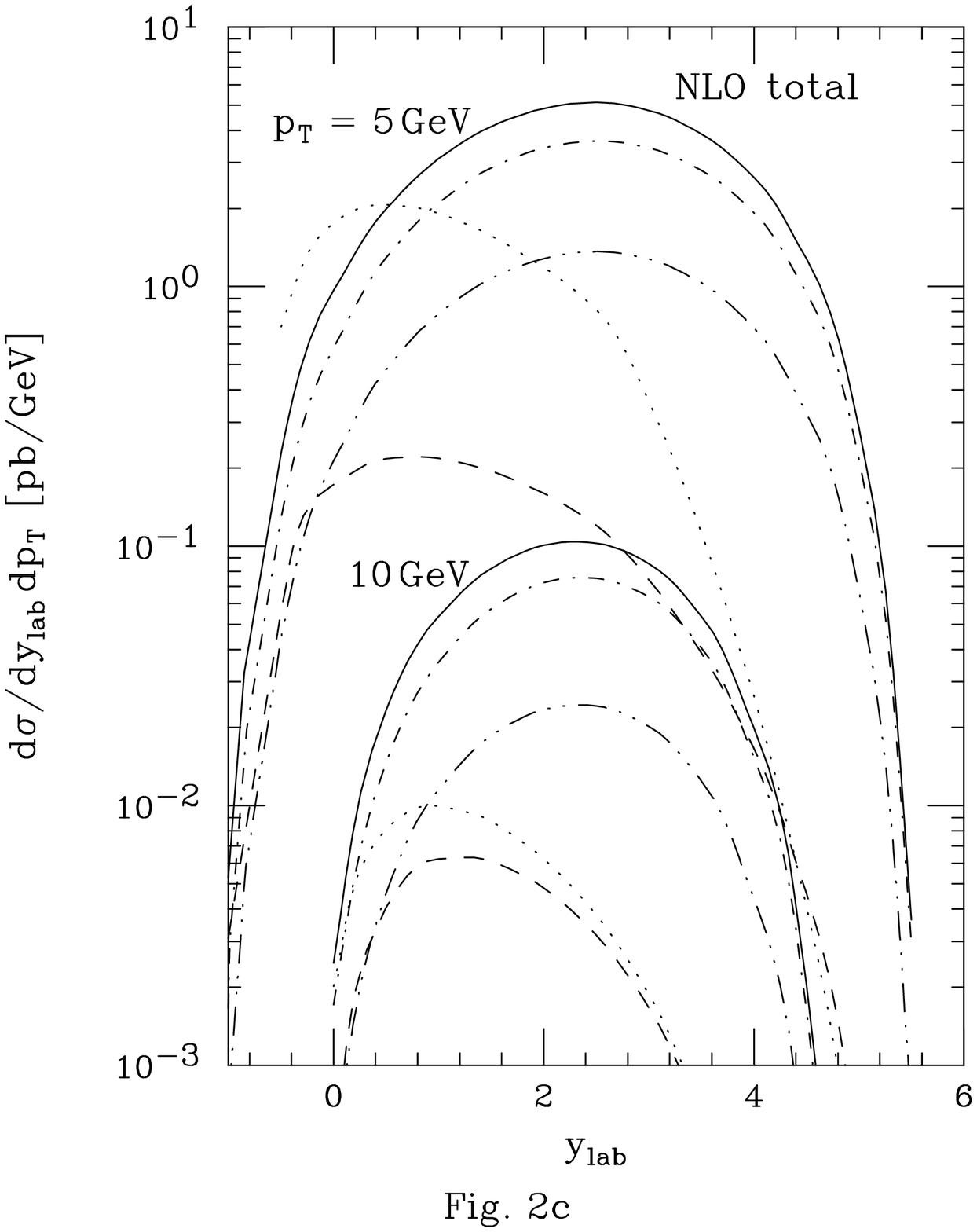,width=\textwidth}
\end{figure}
\newpage
\begin{figure}[ht]
\epsfig{figure=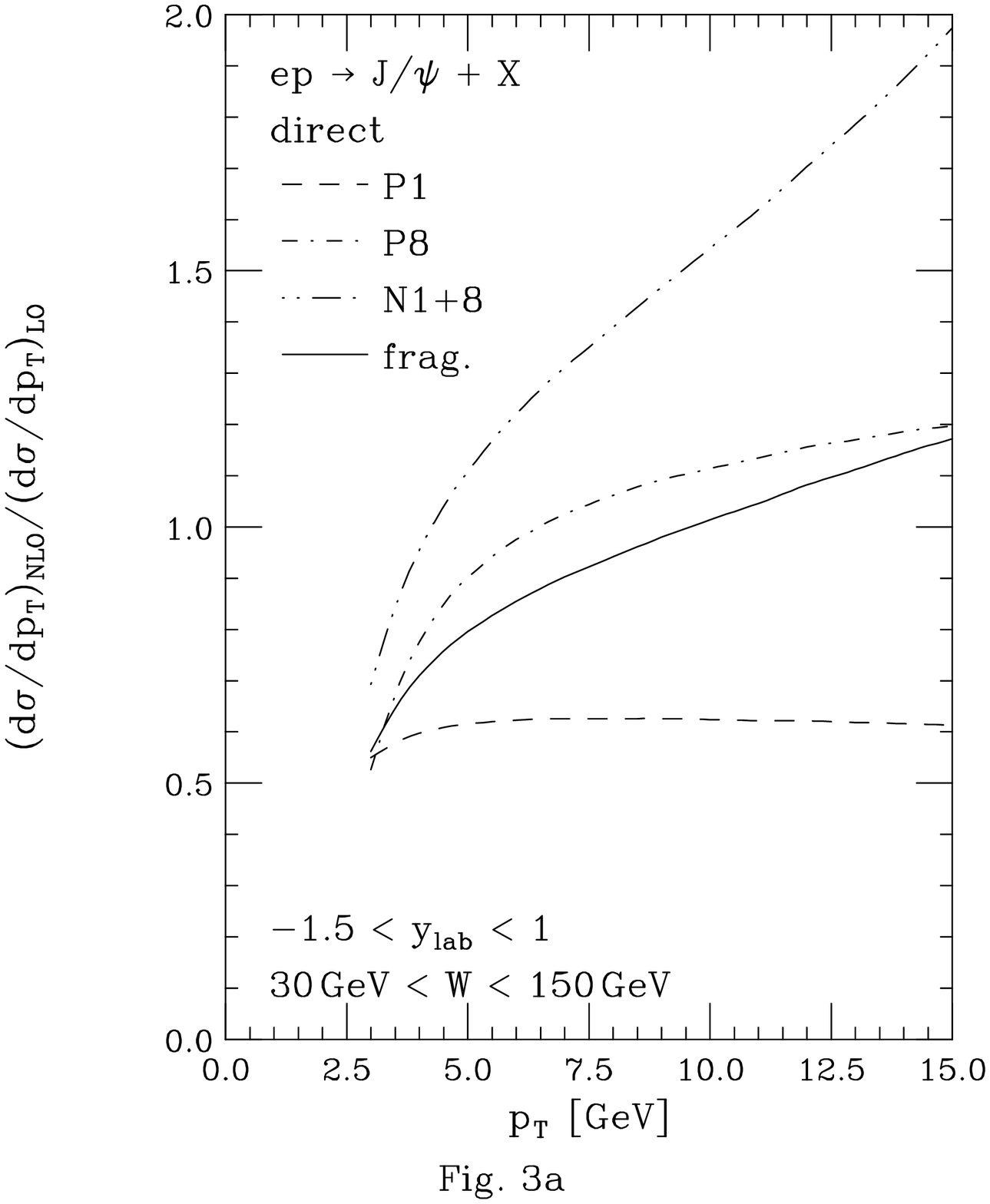,width=\textwidth}
\end{figure}
\newpage
\begin{figure}[ht]
\epsfig{figure=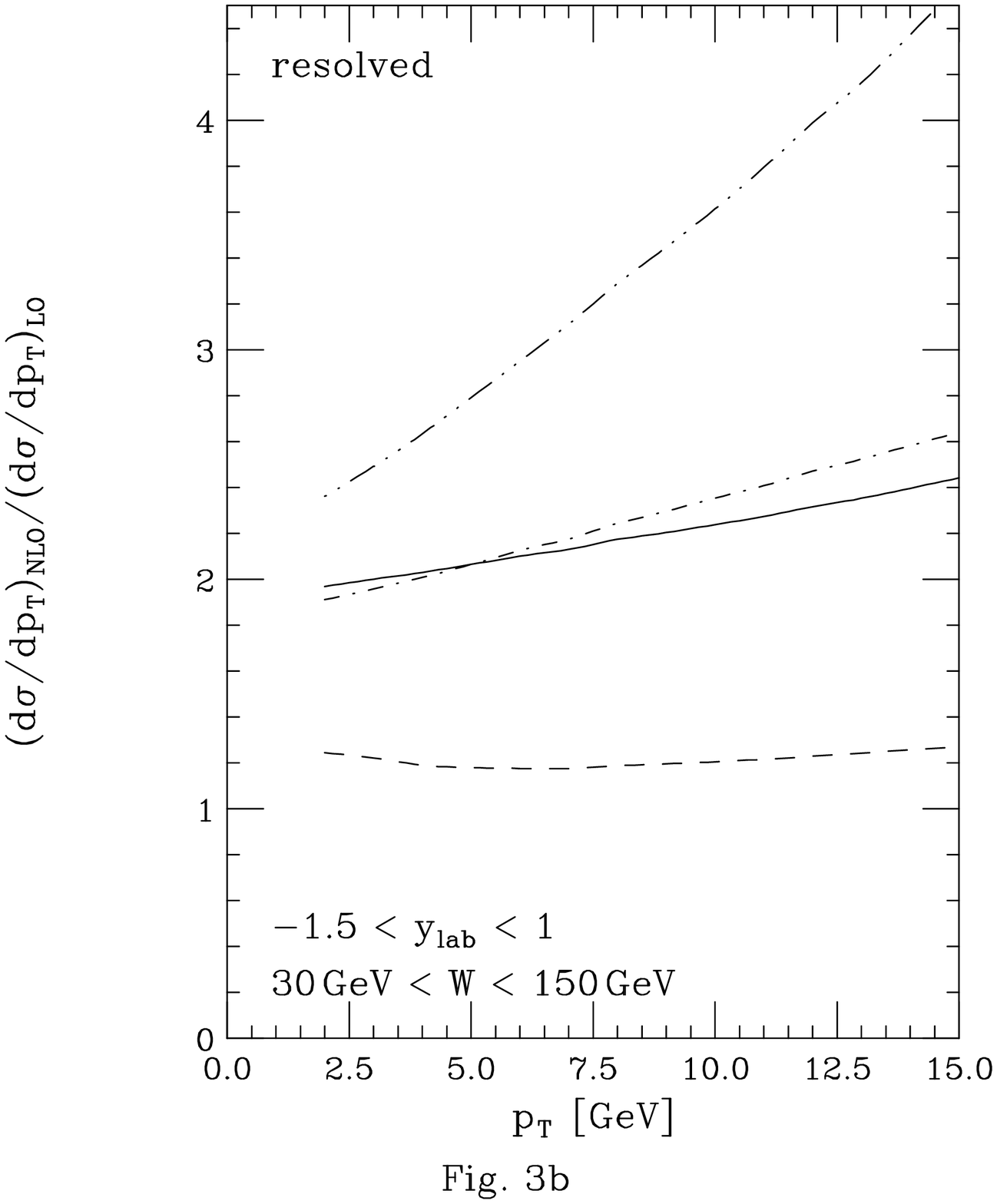,width=\textwidth}
\end{figure}
\newpage
\begin{figure}[ht]
\epsfig{figure=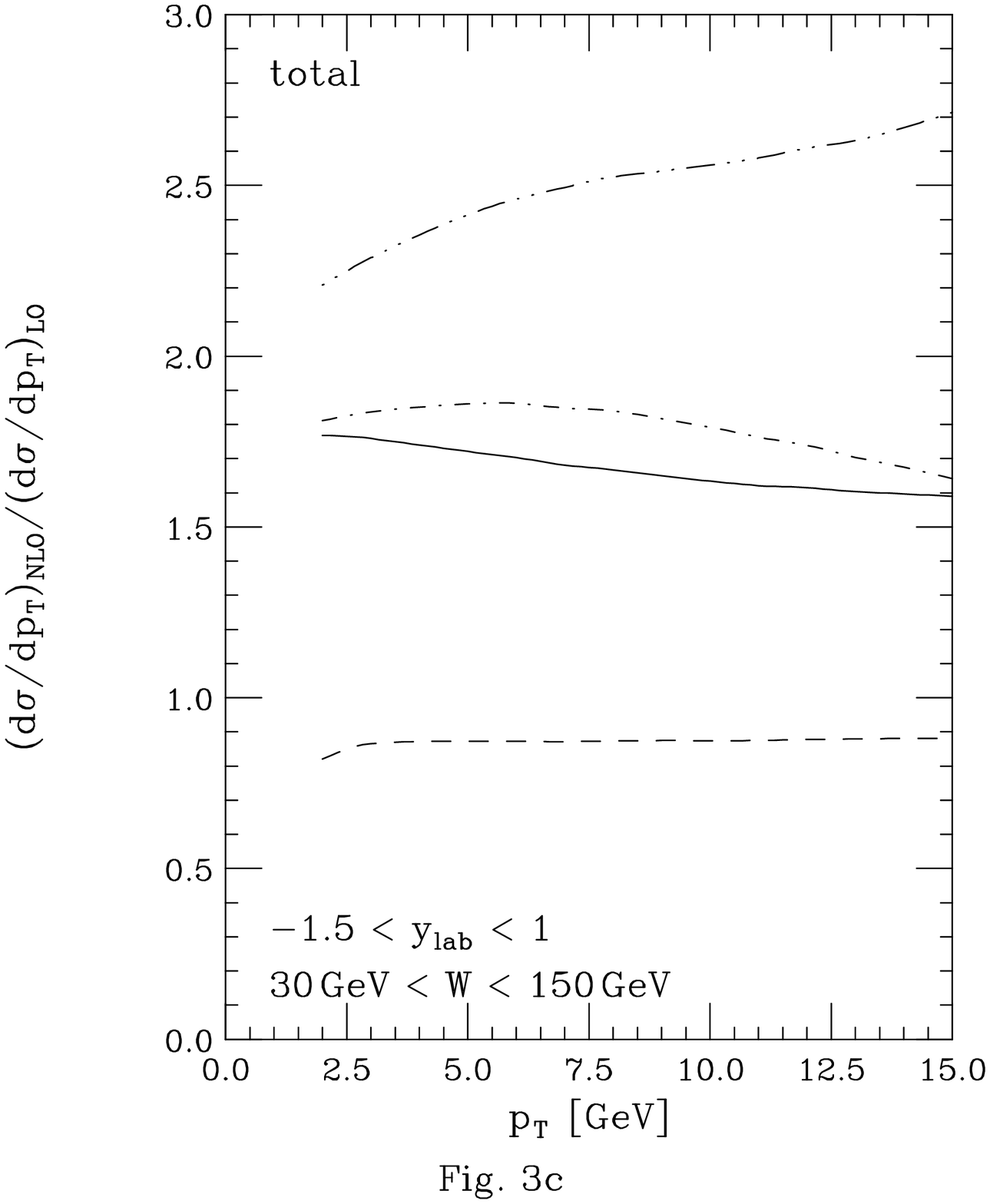,width=\textwidth}
\end{figure}
\newpage
\begin{figure}[ht]
\epsfig{figure=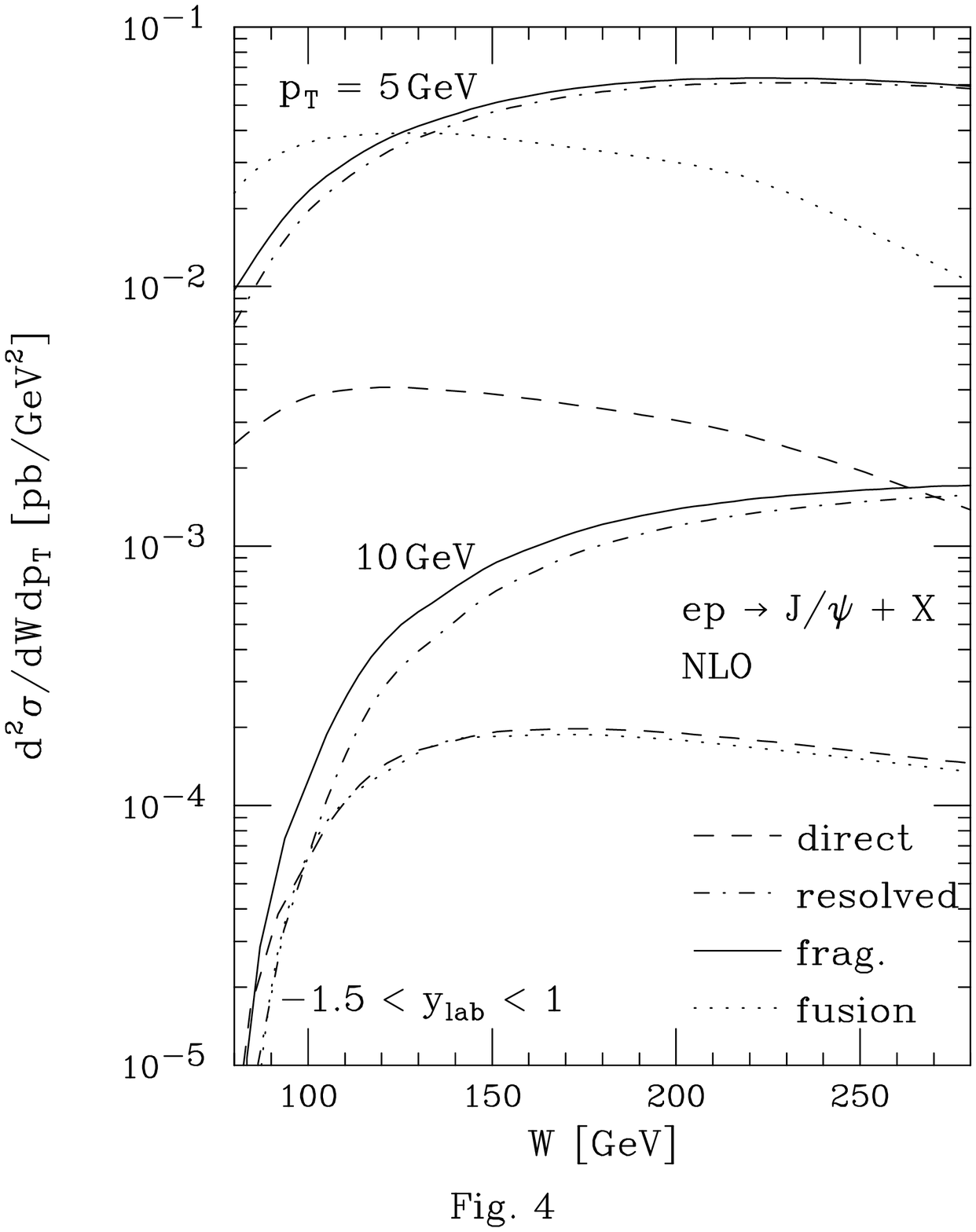,width=\textwidth}
\end{figure}
\newpage
\begin{figure}[ht]
\epsfig{figure=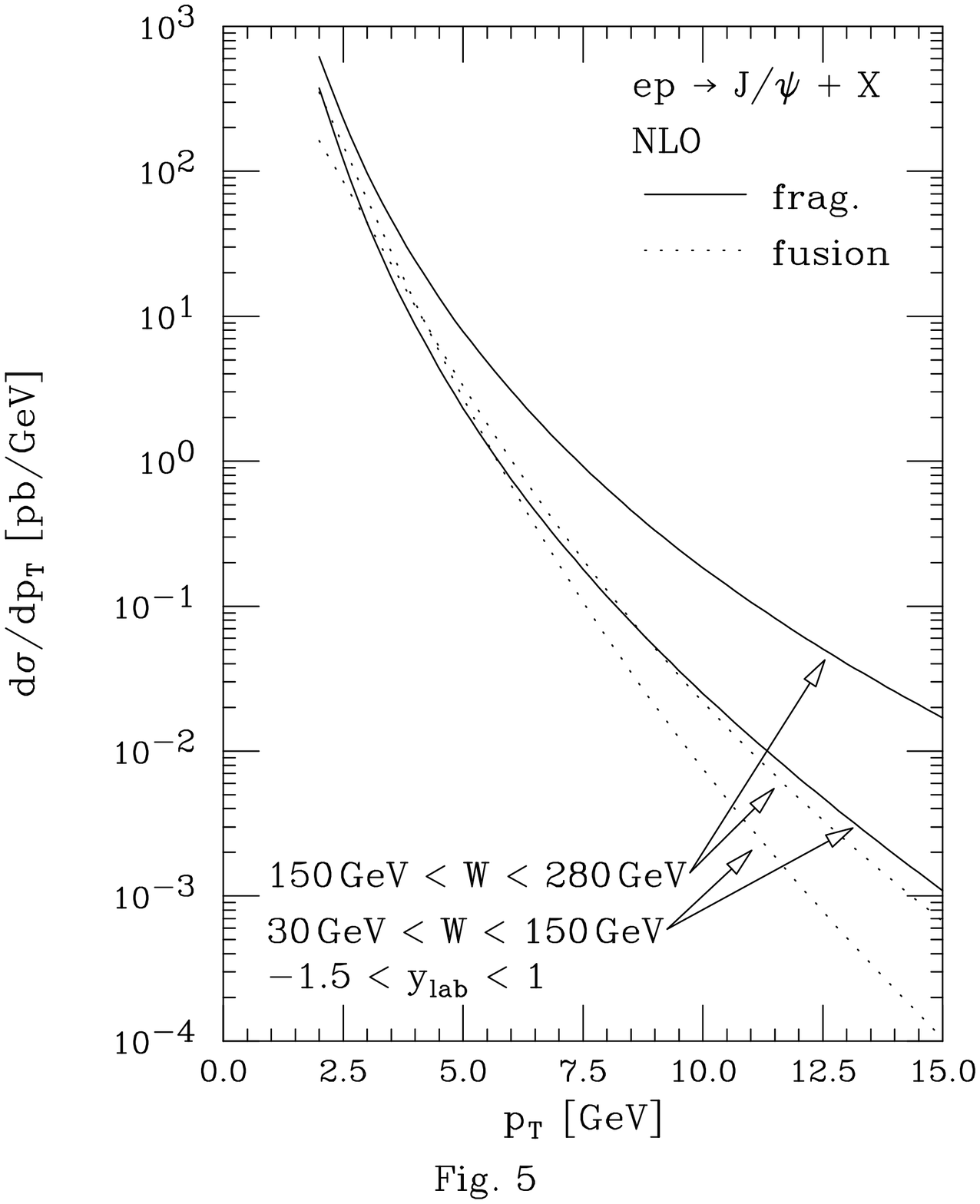,width=\textwidth}
\end{figure}
\newpage
\begin{figure}[ht]
\epsfig{figure=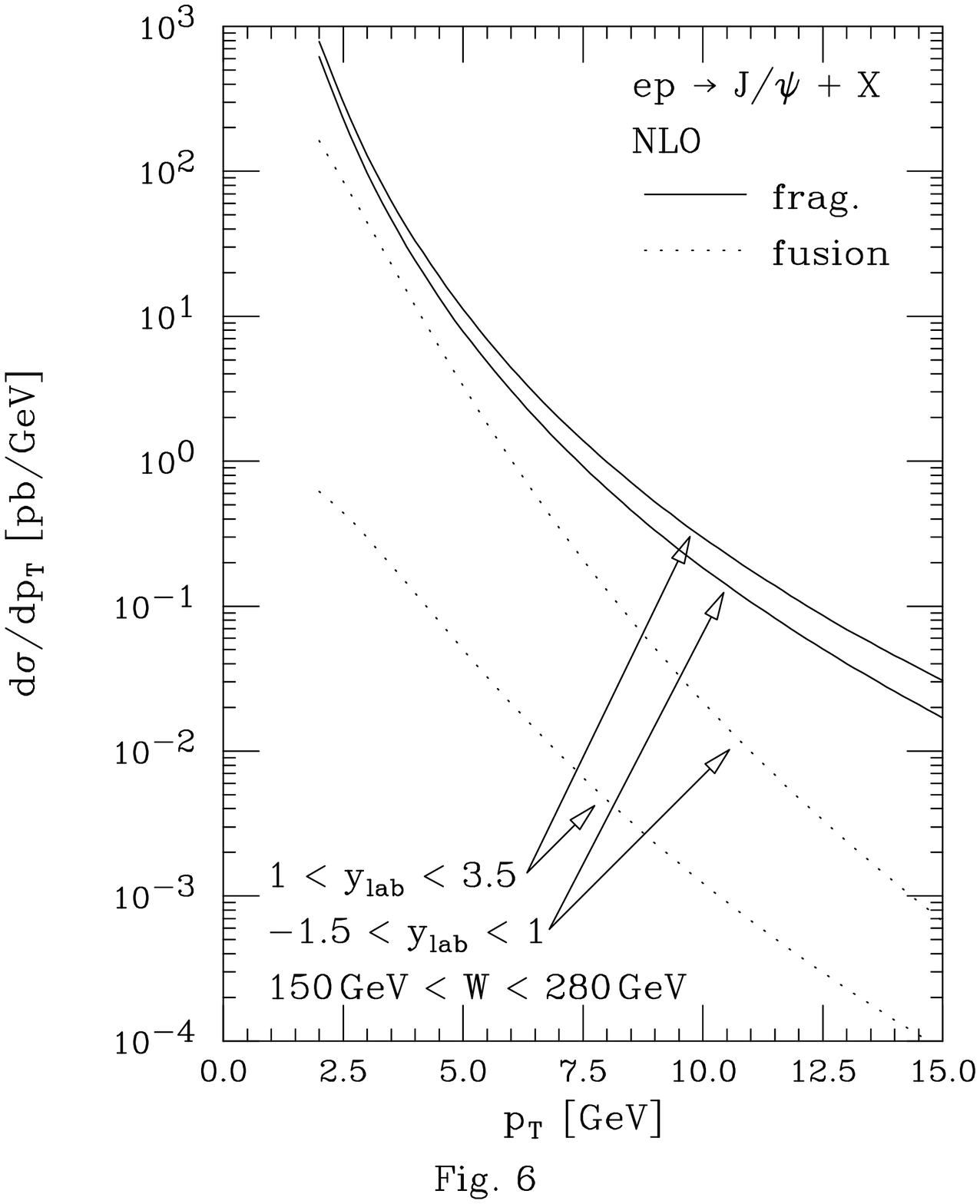,width=\textwidth}
\end{figure}


\begin{thebibliography}{99}

\bibitem{ber} E.L. Berger and D. Jones,
Phys. Rev. D {\bf23}, 1521 (1981);
R. Baier and R. R\"uckl,
Phys.\ Lett.\ {\bf102B}, 364 (1981);
Nucl.\ Phys.\ {\bf B208}, 381 (1982);
{\it ibid.} {\bf B218}, 289 (1983);
J.G. K\"orner, J. Cleymans, M. Kuroda, and G.J. Gounaris,
{\it ibid.} {\bf B204}, 6 (1982);
for a review, see G.A. Schuler, CERN Report Nr.\ CERN--TH.7170/94 
(unpublished).

\bibitem{bai} R. Baier and R. R\"uckl,
Z. Phys.\ C {\bf19}, 251 (1983).

\bibitem{glo} E.W.N. Glover, A.D. Martin, and W.J. Stirling,
Z. Phys.\ C {\bf38}, 473 (1988).

\bibitem{abe} CDF Collaboration, F. Abe {\it et al.},
Phys.\ Rev.\ Lett.\ {\bf69}, 3704 (1992);
{\it ibid.} {\bf71}, 2537 (1993).

\bibitem{bra} E. Braaten and T.C. Yuan,
Phys.\ Rev.\ Lett.\ {\bf71}, 1673 (1993);
Phys.\ Rev.\ D {\bf52}, 6627 (1995);
see also the subsequent papers in Refs.~\cite{don,roy}.

\bibitem{aid} H1 Collaboration, S. Aid {\it et al.},
Nucl.\ Phys.\ {\bf B472}, 3 (1996).

\bibitem{kra} M. Kr\"amer, J. Zunft, J. Steegborn, and P.M. Zerwas,
Phys.\ Lett.\ B {\bf348}, 657 (1995);
M. Kr\"amer,
Nucl.\ Phys.\ {\bf B459}, 3 (1996).

\bibitem{jun} H. Jung, G.A. Schuler, and J.\ Terr\'on,
in {\it Proceedings of the Workshop on Physics at HERA},
Hamburg, Germany, 29--30 October 1991,
edited by W. Buchm\"uller and G. Ingelman, Vol.~2, p.~712;
Int.\ J. Mod.\ Phys.\ A {\bf7}, 7955 (1992).

\bibitem{cho} P. Cho and A.K. Leibovich,
Phys.\ Rev.\ D {\bf53}, 150 (1996);
{\it ibid.} {\bf53}, 6203 (1996).

\bibitem{cac} M. Cacciari and M. Kr\"amer,
Phys.\ Rev.\ Lett.\ {\bf76}, 4128 (1996);
P. Ko, J. Lee, and H.S. Song,
Phys.\ Rev.\ D {\bf54}, 4312 (1996).


\bibitem{mca} M. Cacciari and M. Kr\"amer,
in {\it Proceedings of the Workshop 1995/96 on Future Physics at HERA},
edited by G. Ingelman, A. De Roeck, and R. Klanner, Vol.~1, p.~416.

\bibitem{ebr} E. Braaten and S. Fleming,
Phys.\ Rev.\ Lett.\ {\bf74}, 3327 (1995).

\bibitem{gre} M. Cacciari, M. Greco, M.L. Mangano, and A. Petrelli,
Phys.\ Lett.\ B {\bf356}, 553 (1995).

\bibitem{god} R.M. Godbole, D.P. Roy, and K. Sridhar,
Phys.\ Lett.\ B {\bf373}, 328 (1996);
R.M. Godbole (private communication).

\bibitem{sal} V.A. Saleev,
Mod.\ Phys.\ Lett.\ A {\bf9}, 1083 (1994).

\bibitem{kni} B.A. Kniehl, G. Kramer, and M. Spira,
Report Nos.\ CERN--TH/96--274, DESY~96--210, MPI/PhT/96--103, and
hep--ph/9610267 (October 1996), Z.\ Phys.\ C (in press).

\bibitem{aur} P. Aurenche, R. Baier, A. Douiri, M. Fontannaz, and D. Schiff,
Nucl.\ Phys.\ {\bf B286}, 553 (1987);
F. Aversa, P. Chiappetta, M. Greco, and J.Ph.\ Guillet,
Nucl.\ Phys.\ {\bf B327}, 105 (1989).

\bibitem{bod} G.T. Bodwin, E. Braaten, and G.P. Lepage,
Phys.\ Rev.\ D {\bf51}, 1125 (1995); {\bf55}, 5853(E) (1997).

\bibitem{yua} E. Braaten and T.C. Yuan,
Phys.\ Rev.\ D {\bf50}, 3176 (1994);
E. Braaten and Y.-Q. Chen,
Phys.\ Rev.\ D {\bf55}, 2693 (1997).

\bibitem{che} E. Braaten, K. Cheung, and T.C. Yuan,
Phys.\ Rev.\ D {\bf48}, 4230 (1993);
Y.-Q. Chen, Phys.\ Rev.\ D {\bf48}, 5181 (1993); {\bf50}, 6013(E) (1994).

\bibitem{tcy} T.C. Yuan,
Phys.\ Rev.\ D {\bf50}, 5664 (1994).

\bibitem{bfy} E. Braaten, S. Fleming, and T.C. Yuan,
Annu.\ Rev.\ Nucl.\ Part.\ Sci.\ {\bf46}, 197 (1996).

\bibitem{pdg} Particle Data Group, R.M. Barnett {\it et al.},
Phys.\ Rev.\ D {\bf54}, 1 (1996).

\bibitem{cur} G. Curci, W. Furmanski, and R. Petronzio,
Nucl.\ Phys.\ {\bf B175}, 27 (1980);
W. Furmanski and R. Petronzio,
Phys.\ Lett.\ {\bf97B}, 437 (1980);
P.J. Rijken and W.L. van Neeren,
Nucl.\ Phys.\ {\bf B487}, 233 (1997);
M. Stratmann and W. Vogelsang,
Report Nos.\ DO--TH~96/23, RAL--TR--96--097, and hep--ph/9612250
(December 1996).

\bibitem{don} E. Braaten, M.A. Doncheski, S. Fleming, and M.L. Mangano,
Phys.\ Lett.\ B {\bf333}, 548 (1994).

\bibitem{lon} P. Ernstr\"om, L. L\"onnblad, and M. V\"anttinen,
NORDITA Report Nos.\ NORDITA--96/78~P and hep--ph/9612408 (December 1996).

\bibitem{mel} B. Mele and P. Nason,
Nucl.\ Phys.\ {\bf B361}, 626 (1991);
see also P. Nason, S. Dawson, and R.K. Ellis,
Nucl.\ Phys.\ {\bf B327}, 49 (1989); {\bf B335}, 260(E) (1990).

\bibitem{lai} H.L. Lai, J. Huston, S. Kuhlmann, F. Olness, J. Owens, D. Soper,
W.K. Tung, and H. Weerts,
Phys.\ Rev.\ D {\bf55}, 1280 (1997).

\bibitem{grv} M. Gl\"uck, E. Reya, and A. Vogt,
Phys.\ Rev.\ D {\bf46}, 1973 (1992).

\bibitem{roy} M. Cacciari and M. Greco,
Phys.\ Rev.\ Lett.\ {\bf73}, 1586 (1994);
D.P. Roy and K. Sridhar,
Phys.\ Lett.\ B {\bf339}, 141 (1994).

\bibitem{kar} H1 Collaboration, S. Aid {\it et al.},
Nucl.\ Phys.\ {\bf B472}, 32 (1996);
ZEUS Collaboration,
Contributed Paper No.\ pa05--051 to the 28th International Conference on High
Energy Physics, Warsaw, Poland, 25--31 July 1996.

\bibitem{kie} R. Brugnera, C. Coldewey, C. Kiesling, B. Naroska, and A. Wegner,
private communications.

\bibitem{let} B.A. Kniehl and G. Kramer,
Report Nos.\ DESY~96--036, MPI/PhT/96--018, and
hep--ph/9703280 (March 1997).

E. Braaten and Y.-Q. Chen,
Phys.\ Rev.\ Lett.\ {\bf76}, 730 (1996);

\bibitem{keu} E. Braaten and Y.-Q. Chen,
Phys.\ Rev.\ Lett.\ {\bf76}, 730 (1996);
K. Cheung, W.-Y. Keung, and T.C. Yuan,
Phys.\ Rev.\ Lett.\ {\bf76}, 877 (1996);
F. Yuan, C.-F. Qiao, and K.-T. Chao,
Peking University Report Nos.\ PUTP--96-31 and hep--ph/9701361 (January 1997).

\bibitem{spi} M. Cacciari, M. Greco, B.A. Kniehl, M. Kr\"amer, G. Kramer, and
M. Spira,
Nucl.\ Phys.\ {\bf B466}, 173 (1996).

\end{thebibliography}
\end{document}